\documentclass[journal]{IEEEtran}
\usepackage{cite}
\usepackage{amsmath,amssymb,amsfonts}
\usepackage{algorithmic}
\usepackage{graphicx}
\usepackage{amssymb,comment}
\usepackage{algorithmic}
\usepackage{algorithm}
\usepackage{caption} 
\usepackage{subfig}
\usepackage{subcaption}
\usepackage{scrextend}
\usepackage{verbatim}
\usepackage{color}

\newtheorem{lemma}{\bf Lemma}

\newtheorem{proposition}{\bf Proposition}
\newcommand{\bs}[1]{\boldsymbol{#1}}
\newcommand{\mbf}[1]{\mathbf{#1}}
\newcommand{\ib}[1]{\in\mathbb{#1}}
\newcommand{\ic}[1]{\in\mathcal{#1}}
\newcommand{\ca}[1]{\mathcal{#1}}
\newcommand{\mrm}[1]{\mathrm{#1}}

\usepackage{setspace}
\begin{document}
	\title{Energy-Efficient Movable Antennas: Mechanical Power Modeling and Performance Optimization}
	
	\author{Xin~Wei,~\IEEEmembership{Graduate Student Member,~IEEE,}
		Weidong~Mei,~\IEEEmembership{Member,~IEEE,}
		Xuan~Huang,\\
		Zhi~Chen,~\IEEEmembership{Senior Member,~IEEE,}
		and~Boyu~Ning,~\IEEEmembership{Member,~IEEE}
		\thanks{
			This paper has been presented in part at the 2025 IEEE Global Communications Conference (GLOBECOM) [DOI: 10.1109/GLOBECOM59602.2025.11432508] \cite{wei2025mechanical}. This work was supported in part by the National Key Research and Development Program of China under Grant 2024YFE0200404.
			
			Xin Wei, Weidong Mei, Zhi Chen, and Boyu Ning are with the National Key Laboratory of Wireless Communications, University of Electronic Science and Technology of China, Chengdu 611731, China (e-mail: xinwei@std.uestc.edu.cn, wmei@uestc.edu.cn, chenzhi@uestc.edu.cn, boydning@outlook.com).
			
			Xuan Huang is with the School of Information and Communication Engineering, University of Electronic Science and Technology of China, Chengdu 611731, China (e-mail: huang.xuan@std.uestc.edu.cn).
			}
		}
	
	\maketitle
	\begin{abstract}
		Movable antennas (MAs) offer additional spatial degrees of freedom (DoFs) to enhance wireless communication performance through local antenna movement in a confined region. However, to achieve accurate and fast antenna movement, MA drivers entail non-negligible mechanical power consumption, rendering energy efficiency (EE) optimization more critical compared to conventional fixed-position antenna (FPA) systems. To address this problem, we develop in this paper a fundamental power consumption model for stepper motor-driven multi-MA systems by resorting to basic electric motor theory. Based on this model, we investigate an EE maximization problem for the downlink transmission from a multi-MA base station (BS) to multiple single-antenna users. In particular, we aim to jointly optimize the MAs' positions and moving speeds as well as the BS's transmit precoding matrix subject to collision-avoidance constraints during the multi-MA movements. However, this problem appears to be difficult to be solved optimally. To tackle this challenge, we first reveal that the collision-avoidance constraints can always be relaxed without loss of optimality by properly renumbering the MA indices. For the resulting relaxed problem, we first consider a simplified single-user setup and uncover a hidden monotonicity of the EE performance with respect to the MAs' moving speeds. To solve the remaining optimization problem, we develop a two-layer optimization framework. In the inner layer, the Dinkelbach algorithm is employed to derive the optimal beamforming solution in a semi-closed form for any given MA positions. In the outer layer, a sequential update algorithm is proposed to iteratively refine the MA positions based on the optimal values obtained from the inner layer. Next, we proceed to the general multi-user case and propose an alternating optimization (AO) algorithm to obtain a high-quality suboptimal solution. Numerical results demonstrate that despite the additional mechanical power consumption, the proposed algorithms can outperform both conventional FPA systems and existing EE maximization algorithms that neglect mechanical power consumption.
	\end{abstract}
	\IEEEpeerreviewmaketitle
	
	\begingroup
	\allowdisplaybreaks
	
	\section{Introduction}
	Driven by the growing demand for higher data rates, lower latency, and greater reliability in future sixth-generation (6G) wireless networks, communication system design is increasingly evolving toward architectures that emphasize spectral efficiency, spatial intelligence, and configurational flexibility. Technologies such as massive multiple-input multiple-output (MIMO) \cite{larsson2014massive,yang2021millimeter,ning2021terahertz} and extremely large antenna arrays (ELAAs) \cite{wang2024tutorial} have shown strong potential in exploiting spatial degrees of freedom (DoFs) to achieve significant gains in throughput, interference suppression, and dynamic beamforming. However, these techniques largely rely on fixed-position antennas (FPAs), whose static geometric configurations result in fixed spatial correlations, thus significantly limiting their adaptability to dynamic environments and leading to suboptimal performance \cite{zhu2024movable,zheng2024flexible}.
	
	To address the limitations of FPA architectures, movable antenna (MA) systems (also known as fluid antenna systems \cite{new2025tutorial}) have recently emerged as a promising solution by introducing reconfigurable array geometries \cite{zhu2025tutorial}. Specifically, compared to FPAs, MAs can reposition their elements, either individually or array-wise, by using mechanical actuators or electrically controlled mechanisms within a predefined local region \cite{zhu2024modeling,ning2024movable}. This spatial reconfigurability allows the antenna geometry to dynamically adapt to real-time channel conditions and user locations, thus offering new DoFs for performance enhancement. The promising benefits of MA systems have spurred great enthusiasm in investigating MA position optimization in various wireless systems. For example, the authors in \cite{ma2024mimo} investigated the MA-aided point-to-point MIMO systems and demonstrated that antenna position optimization can significantly enhance MIMO channel capacity by increasing both spatial channel power gains and spatial multiplexing gain. Beyond rate improvement, MAs have also been leveraged to manipulate the spatial correlations among steering vectors, thereby achieving more flexible beam nulling \cite{zhu2023movable,yang2026movable}, multi-beam forming \cite{ma2024multi,liu2025movable}, and wide-beam coverage \cite{wang2025movable,wang2025enhanced}. Furthermore, MA position optimization has been explored in other scenarios such as secure communications \cite{mei2024posistion,tang2025secure,shen2025movable}, single- and multi-user multiple-input single-output (MISO) \cite{mei2024movable,zhu2024enhanced,ma2025robust}, cognitive radio \cite{wei2024joint}, relaying systems \cite{li2025movable}, non-orthogonal multiple access (NOMA) \cite{li2024sum,li2025sum}, intelligent reflecting surface (IRS)-aided wireless communications \cite{wei2025movable,zhang2025sum}, wireless sensing or integrated sensing and communications (ISAC) \cite{ma2024movable,wang2025antenna,lyu2025movable,chen2025antenna}, among others. To further advance the frontier of reconfigurable antenna technologies, novel architectures such as six-dimensional movable antenna (6DMA) \cite{shao2025enhanced,shao2026tutorial,liu2026enabled} and pinching antenna systems \cite{ding2025flexible} have also been proposed recently.
	
	However, despite these prior works, the energy efficiency (EE) of MA systems has not been thoroughly investigated. Particularly, enabling precise position adjustment and motion of MAs typically relies on mechanical actuators (e.g., stepper motors), which can incur non-negligible additional energy consumption. Furthermore, antenna repositioning inevitably incurs a finite movement delay and potential inter-MA collision, reducing the effective duration available for wireless data transmission. Considering these practical factors, it remains unclear whether MAs can still outperform FPAs in terms of EE. To answer this question, a handful of recent works have investigated the EE performance optimization for MA systems \cite{chen2025energy,wu2024globally,ding2025energy}. Specifically, the authors in \cite{chen2025energy} studied the EE maximization for an MA-MIMO system based on statistical channel state information (CSI). However, their analysis considers only the power consumption from data transmission, while ignoring the mechanical energy consumed by antenna movement, which can be substantial in practice. Although the authors in \cite{wu2024globally} incorporate the energy consumption of MA movement into the EE analysis, it models the mechanical power consumption as a fixed constant, overlooking its dependency on motion dynamics such as displacement and speed, and thus lacks a detailed power consumption model for the MA drivers. Recently, a parallel work \cite{ding2025energy} proposed an energy consumption model for mechanically-driven MA systems. However, it adopts a simplified approximation where the mechanical power increases linearly with the moving speed and focuses solely on a single MA at the user side, without theoretically characterizing the impact of antenna velocity on the EE performance or accounting for potential inter-MA collisions in multi-MA movement. In our previous work \cite{wei2025mechanical}, we developed a more practical mechanical power model, but only investigated the EE maximization problem in a simplified single-MA single-user setup.
	
	In this paper, we investigate a more general and more challenging multi-MA multi-user system setup, where a base station (BS) equipped with a linear MA array transmits to multiple single-FPA users at the same time, as shown in Fig.~\ref{Fig_SysModel}. The main contributions of this paper are summarized as follows:
	\begin{enumerate}
		\item We develop an overall power consumption model for the considered multi-MA system that accounts for both the mechanical energy consumed by the MA drivers and the electrical energy consumed by data transmission, by integrating a classical mechanical power consumption model derived from basic electric motor theory. Building on this model, we formulate an EE (i.e., the ratio of the sum rate of all users to the above energy consumption) maximization problem by jointly optimizing the MAs' destination positions and moving speeds as well as the BS's transmit precoding matrix, subject to the collision-avoidance constraints during the multi-MA movements. Although this problem is non-convex and difficult to be optimally solved, we first show that the collision-avoidance constraints can always be relaxed without loss of optimality by properly renumbering the MA indices, thus greatly simplifying the optimization problem.
		\item Next, to solve the resulting relaxed EE maximization problem, we start from the single-user case and reveal a hidden monotonicity of the EE performance with respect to (w.r.t.) the MAs' moving speeds. For the remaining joint BS beamforming and MA position optimization, we recast this problem into a two-layer formulation. In the inner layer, the Dinkelbach algorithm is applied to derive the optimal transmit beamforming in a semi-closed form for any given MA positions. Whereas in the outer layer, the sequential update algorithm is adopted to iteratively optimize the MA positions based on the optimal transmit beamforming derived in the inner layer.
		\item For the general multi-user case, we propose an alternating optimization (AO) algorithm that alternately optimizes the BS precoding and MA positions using the successive convex approximation (SCA) and sequential update methods, respectively. Finally, simulation results demonstrate that our proposed algorithms consistently outperform both conventional FPA-based systems and existing EE maximization schemes that ignore mechanical power consumption, under both single-user and multi-user setups. It is also shown that the optimized MA positions by our proposed algorithm achieve a superior balance between rate improvement via antenna movement and the associated mechanical power consumption compared to other baseline schemes.
	\end{enumerate}
	
	The rest of this paper is organized as follows. Section II presents the overall power consumption model for our considered mechanically-driven MA system and formulates the EE maximization problem. Sections III and IV present our proposed solutions to this problem in the single- and multi-user cases, respectively. Section V presents the numerical results. Finally, Section VI concludes this paper and discusses future directions.
	
	\begin{figure}[!t]
		\centering
		\captionsetup{justification=raggedright,singlelinecheck=false}
		\centerline{\includegraphics[width=0.4\textwidth]{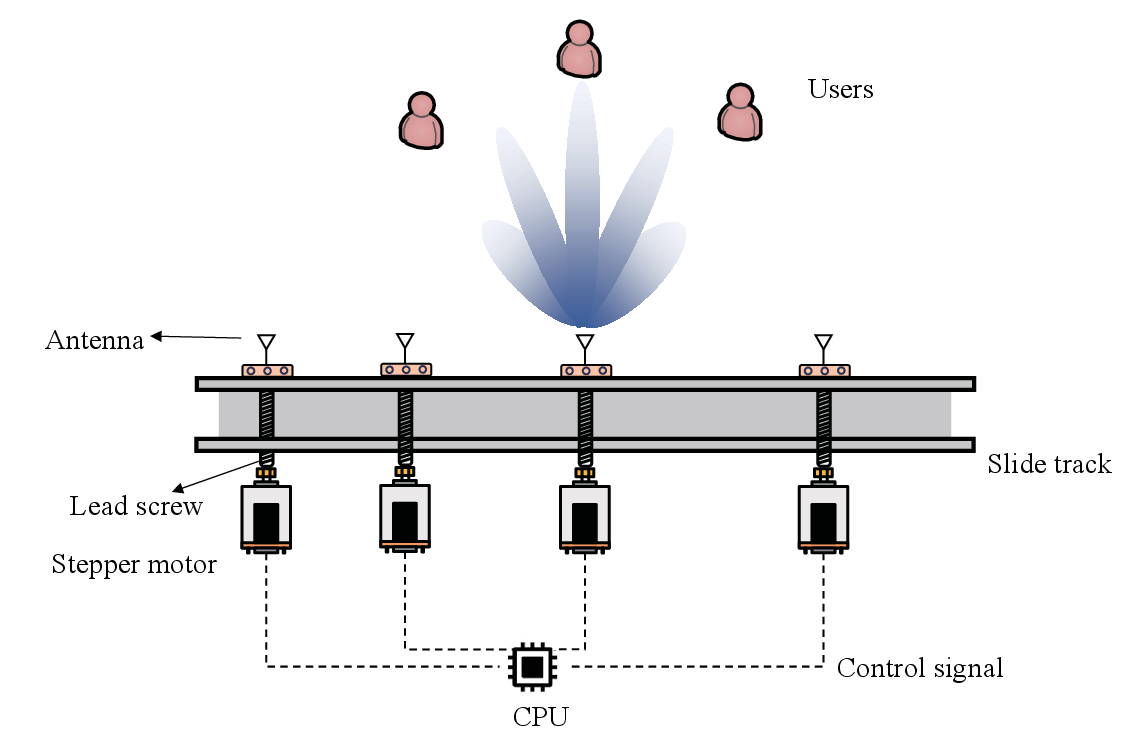}}
		\captionsetup{font=footnotesize}
		\caption{Stepper motor-driven MA system with multiple users.}
		\label{Fig_SysModel}
		\vspace{-15pt}
	\end{figure}
	
	{\it Notations:} $a$, $\bs{a}$, $\mbf{A}$, and $\ca{A}$ denote a scalar, a vector, a matrix and a set, respectively. For a complex number $a$, $\angle a$, $|a|$, and $a^*$ denote its phase, amplitude and conjugate, respectively. $(\cdot)^T$, $(\cdot)^H$, and $(\cdot)^{-1}$ denote the transpose, conjugate transpose and inverse of a matrix, respectively. $\mathbb{R}$ and $\mathbb{C}$ denote the sets of real numbers and complex numbers, respectively. $|a|$ and $||\bs{a}||_2$ denote the amplitude of a scalar $a$ and the norm of a vector $\bs{a}$, respectively. $\mbf{A}[n,:]$ and $\mbf{A}[:,m]$ denotes the $n$-th row and the $m$-th column of a matrix $\mbf{A}$, respectively. $x\sim\ca{CN}(\mu,\sigma^2)$ represents that $x$ follows the circularly symmetric complex Gaussian (CSCG) distribution with the mean $\mu$ and variance $\sigma^2$. $\mrm{sign}(a)$ is the sign function which is given by $\mrm{sign}(a)\triangleq\begin{cases}
		\frac{a}{|a|},\,\quad a\ne 0\\
		0,\qquad a=0
	\end{cases}$.
	
	\begingroup
	\allowdisplaybreaks
	\section{System Model and Problem Formulation}
	\subsection{System Model}
	\begin{figure}[!t]
		\centering
		\captionsetup{justification=raggedright,singlelinecheck=false}
		\centerline{\includegraphics[width=0.3\textwidth]{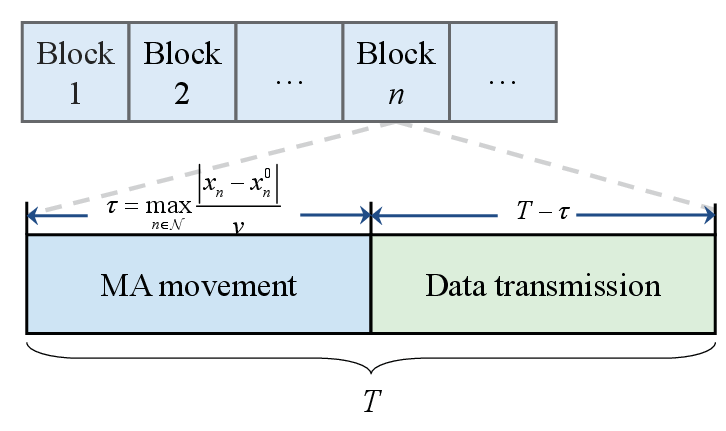}}
		\captionsetup{font=footnotesize}
		\caption{Two-stage transmission protocol in mechanically-driven MA systems.}
		\label{Fig_Protocol}
		\vspace{-21pt}
	\end{figure}
	As shown in Fig. 1, we consider an MA-enhanced communication system where a BS equipped with $N$ MAs serves $K$ users each with a single FPA. We consider that the positions of the $N$ MAs can be flexibly adjusted within a linear array with the length of $A$ meters (m). Each MA is assumed to be mechanically driven via a dedicated stepper motor, which employs a lead screw with an outer radius of $l_0$ as its linear actuator. Considering the discrete nature of stepper motors which rotate by a fixed step angle, the MAs can only move in a discrete step size rather than continuously. Let $\omega_{D}$ denote the step angle of the stepper motor. Then, the corresponding step size of each MA is expressed as $d_s=\omega_{D}l_0$. As a result, the continuous linear array is discretized into $M=\lfloor \frac{A}{d_s}\rfloor$ points as the MAs' candidate positions, where $\lfloor\cdot\rfloor$ denotes the round-down operation. Let $\ca{C}_t\triangleq\left\{0,d_s,2d_s,\cdots,(M-1)d_s\right\}$ denote the set of all candidate positions and $x_n^0$, $x_n^0\ic{C}_t$ denote the initial position of the $n$-th MA, $n\ic{N}\triangleq\left\{1,2,\cdots,N\right\}$. Define $\bs{x}^0\triangleq[x_1^0,x_2^0,\cdots,x_N^0]^T\ib{R}^{N\times1}$ as the current position vector (CPV) of all MAs. Without loss of generality, we assume $x_1^0<x_2^0<\cdots<x_N^0$.
	
	We assume a block-fading channel model and focus on a given time block of duration $T$, which is divided into two stages, as illustrated in Fig.~\ref{Fig_Protocol}. Specifically, in the first stage, each stepper motor drives its associated MA to its destination position. We assume that the BS remains nearly inactive in this phase with negligible energy consumption; hence, the power consumption in this stage is mainly from the stepper motors. Let $x_n\ic{C}_t$, $n\ic{N}$ denote the destination position of the $n$-th MA and $\bs{x}=\left[x_1,x_2,\cdots,x_N\right]^T\ib{R}^{N\times1}$ denote the destination positions vector (DPV) of all MAs. To reduce the modeling and controlling complexity, we assume that all MAs move at the same constant speed, denoted by $v$. As such, the delay for the $n$-th MA to move from its current position to the destination position is given by
	\begin{equation}
		\tau_n\triangleq\frac{\left|x_n-x_n^0\right|}{v},\,\,\forall n\ic{N}.
	\end{equation}
	Moreover, to avoid mutual coupling among all MAs, we set
	\begin{equation}\label{eqn_MinDist}
		\left|x_i-x_j\right|\ge D_{\min},\,\,\forall i,j\ic{N},\,\,i\ne j,
	\end{equation}
	where $D_{\min}$ is the minimum inter-antenna spacing to avoid mutual coupling.
	
	The first stage ends after all MAs reach their associated destination positions. Hence, the actual time delay in antenna movement is given by
	\begin{equation}
		\tau\triangleq\underset{n\ic{N}}{\max}\,\,\tau_n.
	\end{equation}
	To ensure that the antenna movement can be finished within the channel coherence time, it must hold that $\tau < T$, or equivalently,
	\begin{equation}\label{eqn_Velocity}
		v\ge\underset{n\ic{N}}{\max}\,\,\frac{|x_n-x_n^0|}{T}.
	\end{equation}
	
	It is also worth noting that there may exist inter-MA collisions in moving the MAs from their current positions to the destination positions, as illustrated in Fig.~\ref{Fig_Collision}(a). To prevent such collisions, the distance between any two distinct MAs must remain greater than a prescribed threshold, denoted by $D_{\mrm{th}}$ (with $D_{\mrm{th}}\ge D_{\min}$), at all times during the movement. Specifically, given the current and destination positions of the $n$-th MA, i.e., $x_n^0$ and $x_n$, its position at any time instant within the movement period is given by
	\begin{equation}
		s_n(t)\triangleq\begin{cases}
			x_n^0+vt\cdot\mrm{sign}\left(x_n-x_n^0\right),\quad 0 \le t < \tau_n,\\
			x_n,\qquad\qquad\,\,\qquad\qquad\qquad \tau_n \le t \le \tau.
		\end{cases}
	\end{equation}
	Consequently, the collision-avoidance constraint is given by
	\begin{equation}\label{eqn_Collision}
		\left|s_i(t)-s_j(t)\right|\ge D_{\mrm{th}},\,\,\forall i,j\ic{N},i\ne j,\,\,t \in [0,\tau].
	\end{equation}\vspace{-30pt}
	
	\subsection{Power Consumption Model}
	\begin{figure}[!t]
		\centering
		\captionsetup{justification=raggedright,singlelinecheck=false}
		\centerline{\includegraphics[width=0.45\textwidth]{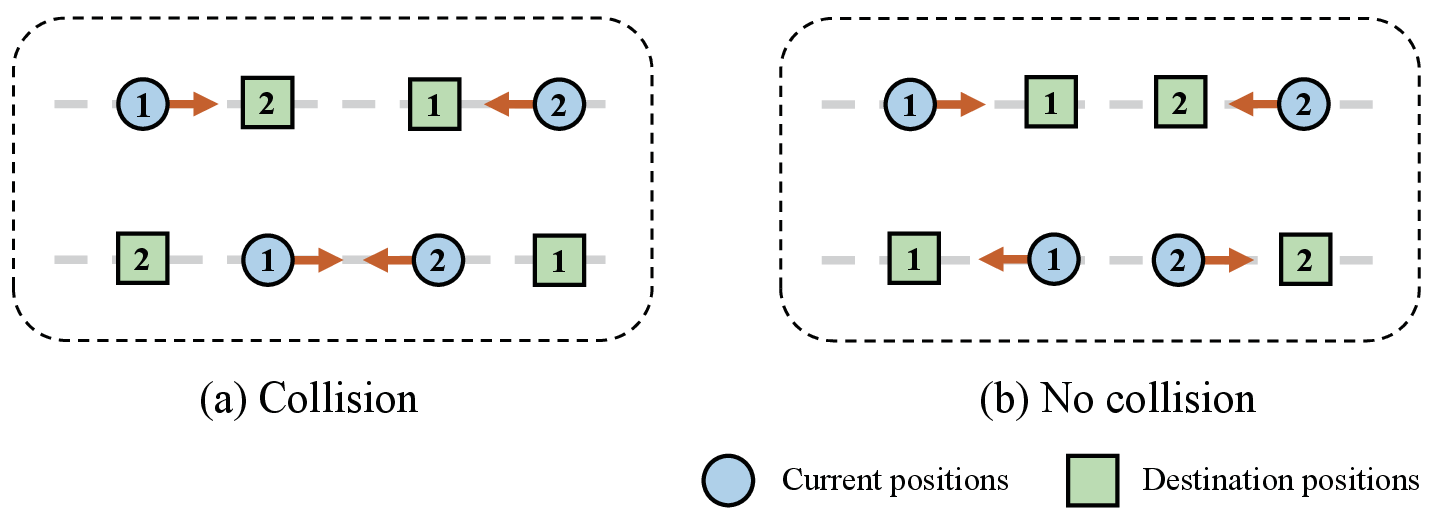}}
		\captionsetup{font=footnotesize}
		\caption{Illustration of multi-MA movements with $N=2$.}
		\label{Fig_Collision}
		\vspace{-18pt}
	\end{figure}
	The total power consumption of the considered MA system comprises the energy consumed by the MA drivers in the first stage and that radiated for data transmission in the second stage, i.e.,\footnote{For simplicity, we assume that all stepper motors are identical and share the same torque–speed characteristics. Nevertheless, the proposed power consumption model remains applicable to scenarios with different stepper motors, as their varying characteristics can be captured by simply updating the specific physical parameters (e.g., resistance and inductance) in the developed power consumption model.}
	\begin{equation}\label{eqn_PwCsp}
		E_{\mrm{total}}\left(\mbf{W},\bs{x},v\right)=\underset{\text{First stage}}{\underbrace{\sum_{n=1}^{N}{\tau_n P_M\left(v\right)}}}+\underset{\text{Second stage}}{\underbrace{\left(T-\tau\right)P_D\left(\mbf{W}\right)}},\vspace{-5pt}
	\end{equation}
	where $P_M\left(v\right)$ and $P_D\left(\mbf{W}\right)$ denote the power consumption for the MA driver and data transmission, respectively. The matrix $\mbf{W}\triangleq\left[\bs{w}_1,\bs{w}_2,\cdots,\bs{w}_K\right]\ib{C}^{N\times K}$ represents the transmit precoding matrix for the $K$ users with $\bs{w}_k\ib{C}^{N\times1}$ denoting the transmit beamforming vector for the $k$-th user. For the power consumption model during data transmission, it primarily includes the radiated power used for signal transmission and the static circuit power consumption, which is given by \cite{xu2013energy}
	\begin{equation}
		P_{D}\left(\mbf{W}\right)=\frac{1}{\zeta}\mathrm{Tr}\left(\mbf{W}\mbf{W}^H\right)+P_{s},
	\end{equation}
	where $P_{s}$ denotes the static circuit power consumption, and $\zeta\in(0,1]$ denotes the efficiency of the power amplifier (PA). Here, we assume an ideal PA (i.e., $\zeta=1$) in the following analysis for ease of exposition, as $\zeta$ only acts as a scaling factor without affecting the efficacy of our proposed algorithms.
	
	Next, according to basic electric motor theory, the power consumption of the stepper motor primarily results from the mechanical work required to drive the load, which depends on the MA's moving speed $v$ and is given by \cite{acarnley2002stepping}
	\begin{equation}\label{eqn_MotorPower}
		P_M\left(v\right)=\omega M(\omega)=\frac{v}{l_0}M\left(\frac{v}{l_0}\right),
	\end{equation}
	where $\omega=v/l_0$ denotes the angular velocity of the stepper motor associated with each MA. In addition, $M(\omega)$ represents the pull-out torque of the stepper motor and is characterized as \cite{acarnley2002stepping}
	\begin{equation}\label{eqn_Torque}
		M(\omega)=\frac{p\psi_{\mrm{M}}V}{\sqrt{R^2+\omega^2L^2}}-\frac{p\omega\psi_{\mrm{M}}^2R}{R^2+\omega^2L^2},
	\end{equation}
	where $p$ denotes the number of the rotor teeth, $\psi_{\mrm{M}}$ denotes the peak magnet flux linking each winding, $V$ denotes the voltage, $R$ denotes the phase resistance, and $L$ denotes the phase induction.
	
	\begin{figure}[!t]
		\centering
		\subfloat[]{
			\includegraphics[width=0.5\linewidth]{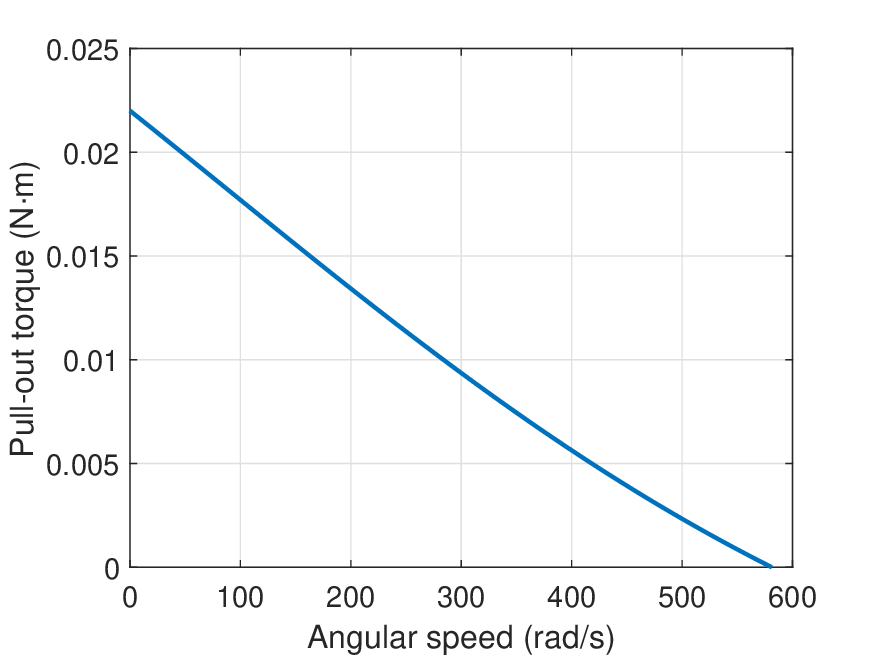}
		}
		\hfil
		\hspace{-20pt}
		\subfloat[]{
			\includegraphics[width=0.5\linewidth]{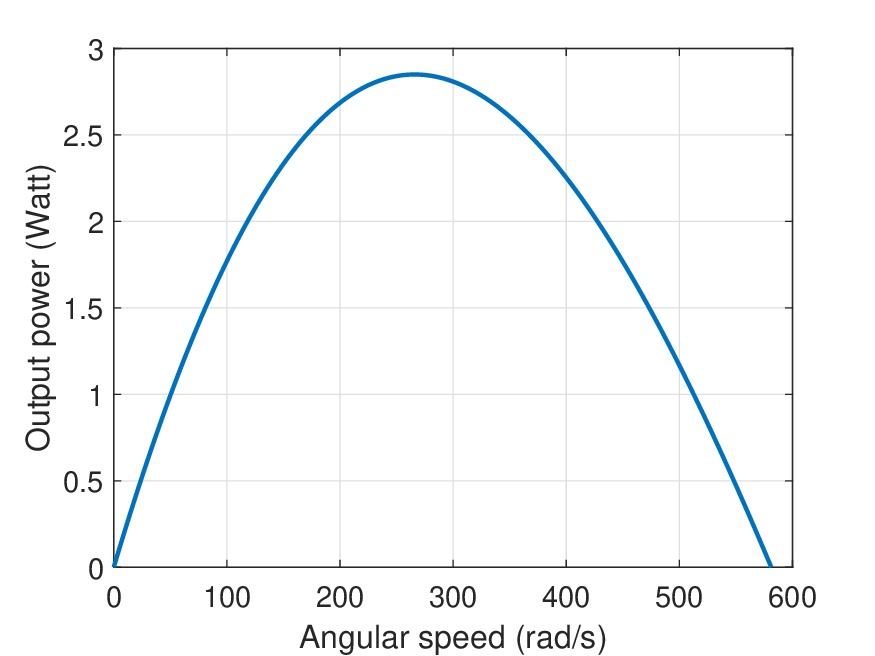}
		}
		\captionsetup{font=footnotesize}
		\caption{(a) Pull-out torque; (b) Output power versus the angular speed of the stepper motor.}
		\vspace{-20pt}
		\label{Fig_Motor}
	\end{figure}
	
	To facilitate a deeper understanding of the mechanical power consumption, we consider a simplified case where a single stepper motor drives a single MA, and plot in Figs.~\ref{Fig_Motor}(a) and \ref{Fig_Motor}(b) the pull-out torque in \eqref{eqn_Torque} and the power consumption in \eqref{eqn_MotorPower} versus the speed of its driven load, i.e., $v$, respectively. The voltage is $V=11.94$ volt (V), the phase resistance is $R=75$ Ohm ($\Omega$), the phase inductance is $L=65.6$ millihenry (mH), the number of rotor teeth is $p=6$, the peak magnet flux linking each winding is $\psi_{M}=0.023$ Weber (Wb), and the radius of the lead screw is $l_0=5$ millimeter (mm).
	
	It is observed from Fig.~\ref{Fig_Motor}(a) that the pull-out torque of the stepper motor gradually decreases with its angular velocity $\omega$. This is because the pull-out torque produced by a stepper motor is proportional to the current through the coils. As $\omega$ increases, the back electromotive force (EMF) generated by the motor's windings increases, which acts against the applied voltage and reduces the effective voltage driving the current through the motor coils. On the other hand, stepper motors are inductive loads. The inductive resistance of the coils also increases with $\omega$, which limits the current that can be supplied to the windings. Hence, it is shown in Fig.~\ref{Fig_Motor}(b) that the power consumption of a stepper motor first increases with $v$, achieving its maximum value, and then decreases, rather than remaining constant. It is also worth noting that the power consumption drops to zero when the angular speed reaches its maximum value (denoted by $\omega_{M}$), corresponding to the no-load condition, which is typically unachievable in practice due to the non-negligible weight of the MA. Therefore, we set a maximum achievable angular speed $\omega_{\max}$, $\omega_{\max}<\omega_{M}$ for the stepper motor in this paper, accounting for the load of the MA. As such, the maximum moving speed of the MA is given by $v_{\max}=\omega_{\max}l_0$.
	
	\textbf{Remark 1}: It is worth noting that the proposed power consumption model is specifically tailored for stepper motor-driven MAs, differing fundamentally from other architectures like electronically driven MAs \cite{ning2024movable}. This is because the power consumption of the latter type of MAs is governed by static switching circuits or fluid dynamics, rather than the characteristics of stepper motors modeled in \eqref{eqn_MotorPower} and \eqref{eqn_Torque}. Moreover, the collision-avoidance constraints in \eqref{eqn_Collision} are unique to our considered system and typically absent for electrically driven MAs.\vspace{-10pt}
	
	\subsection{Problem Formulation}
	Based on \eqref{eqn_Collision}-\eqref{eqn_Torque}, we formulate an EE maximization problem for the considered MA system in this subsection. Let $\bs{h}_k\left(\bs{x}\right)\triangleq\left[h_k(x_1),h_k(x_2),\cdots,h_k(x_n)\right]\ib{C}^{N\times1}$ denote the BS-user $k$ channel w.r.t. DPV $\bs{x}$. Then, the received signal at user $k$ is given by
	\begin{equation}
		y_k=\underset{\text{Desired signal}}{\underbrace{\bs{h}_k^H(\bs{x})\bs{w}_k s_k}}+\underset{\text{Interference signal}}{\underbrace{\sum_{i\ne k}{\bs{h}_k^H(\bs{x})\bs{w}_i s_i}}}+n_k,\forall k\ic{K},
	\end{equation}
	where $s_k$ denotes the transmitted data symbol for user $k$ satisfying $\mathbb{E}[|s_k|^2]=1$, $\forall k\ic{K}$ and $n_k$ represents the received noise at user $k$ with $n_k\sim\ca{CN}(0,\sigma^2)$, where $\sigma^2$ denotes the average noise power at each user. It is also assumed that the transmitted symbols for different users are uncorrelated, such that $\mathbb{E}[s_is_j^*]=0$, $\forall i,j\ic{K}, i\ne j$. Hence, the signal-to-interference-and-noise ratio (SINR) at user $k$ is given by
	\begin{equation}
		\gamma_k\left(\mbf{W},\bs{x}\right)=\frac{\left|\bs{h}_k^H\left(\bs{x}\right)\bs{w}_k\right|^2}{\sum_{i\ne k}^{K}{\left|\bs{h}_k^H\left(\bs{x}\right)\bs{w}_i\right|^2}+\sigma^2},\,\,\forall k\ic{K}.
	\end{equation}
	Accordingly, the achievable rate of user $k$ is given by
	\begin{equation}\label{eqn_Rate}
		R_k\left(\mbf{W},\bs{x}\right)=\log_2\left(1+\gamma_k\left(\mbf{W},\bs{x}\right)\right).
	\end{equation}
	As such, the energy efficiency of the considered multi-MA system within a channel coherence block can be expressed as
	\begin{equation}\label{eqn_EE}
		\mrm{EE}(\mbf{W},\bs{x},v)=\frac{(T-\tau)\sum_{k=1}^{K}R_k\left(\mbf{W},\bs{x}\right)}{E_{\mrm{total}}\left(\mbf{W},\bs{x},v\right)}.
	\end{equation}
	It is observed from \eqref{eqn_EE} that the EE performance depends on the speed of the MA, $v$. As $v$ increases, the movement delay $\tau$ will decrease, which helps increase the achievable rate in the numerator of \eqref{eqn_EE}. However, it remains unclear whether the total power consumption in the denominator of \eqref{eqn_EE} will increase or not. As such, there exists a non-trivial relationship between the EE in \eqref{eqn_EE} and the MA's speed $v$. Furthermore, the EE in \eqref{eqn_EE} also depends on the DPV $\bs{x}$, as it affects both the communication channel quality and the movement delay for a given $v$. Last but not least, the BS's transmit precoding matrix $\mbf{W}$ can also affect \eqref{eqn_EE} by increasing the achievable rates of the users via beamforming optimization.
	
	Hence, in this paper, we aim to maximize \eqref{eqn_EE} by jointly optimizing the DPV $\bs{x}$, the BS's transmit precoding matrix $\mbf{W}$, and the moving speed of the MA $v$. The associated optimization problem can be formulated as
	\begin{subequations}
		\begin{align}
			(\text{P1})\quad\underset{\mbf{W},\bs{x},v}{\max}\quad&\mrm{EE}(\bs{x},\mbf{W},v)\nonumber\\
			\mrm{s.t.}\quad & \mrm{Tr}\left(\mbf{W}\mbf{W}^H\right) \le P_{\max},	\label{eqn_C1}\\
			& x_n\ic{C}_t,\,\,\forall n\ic{N}, \label{eqn_C2}\\
			& \underset{n\ic{N}}{\max}\,\,\frac{|x_n-x_n^0|}{T} \le v \le v_{\max}.\label{eqn_C3}\\
			& \eqref{eqn_MinDist},\eqref{eqn_Collision}.\nonumber
		\end{align}
	\end{subequations}
	Note that for (P1), if the optimized DPV $\bs{x}$ is set identical to the CPV, i.e., $x_n=x_n^0$, $\forall n\ic{N}$, the MA system reduces to the conventional FPA system. As such, the optimal EE value of (P1) is ensured to be no worse than that by an FPA system. Moreover, if the channel coherence time $T$ is sufficiently large, e.g., $T \rightarrow \infty$, the EE in \eqref{eqn_EE} will degrade to conventional EE without accounting for the mechanical power consumption, i.e.,
	\begin{equation}\label{eqn_EE_infty}
		\mrm{EE}(\mbf{W},\bs{x},v)=\frac{\sum_{k=1}^{K}R_k\left(\mbf{W},\bs{x}\right)}{\mrm{Tr}\left(\mbf{W}\mbf{W}^H\right)+P_{s}},
	\end{equation}
	as studied in \cite{chen2025energy} and \cite{wu2024globally}. It follows that a different DPV from that in \cite{chen2025energy} and \cite{wu2024globally} is generally needed considering the mechanical power consumption, especially if the channel coherence time is not long.
	
	\textbf{Remark 2}: It is worth noting that to characterize the fundamental limit of the EE of the considered MA system, we assume that all required channel state information (CSI) is available via the existing channel estimation techniques for MA systems; see e.g., \cite{ma2023compressed,xiao2024channel,zhang2024channel,huang2025cnn}. In particular, in the considered multi-MA setup, antenna movement is not strictly necessary for channel estimation, especially for sparse channels, since existing channel estimation techniques for FPAs can be directly applied to recover the channel parameters. This, in turn, avoids additional mechanical power consumption and movement-induced delay.
	
	\textbf{Remark 3:} Note that in (P1), although no explicit constraint is imposed on the data transmission time, the structure of the objective function itself inherently prevents the solution from falling into an arbitrarily short transmission interval. Furthermore, our proposed algorithms and theoretical results still hold in the presence of a minimum transmission-time constraint, since this only increases the lower bound of the antenna movement velocity in \eqref{eqn_C3} and therefore does not affect the effectiveness of the proposed algorithms.
	
	However, (P1) is difficult to be solved optimally due to the collision-free constraints in \eqref{eqn_Collision}, the fractional form of the objective function, and the nonlinear and intricate expression of $P_M(v)$. Fortunately, it can be shown that we can safely drop \eqref{eqn_Collision} without loss of optimality of (P1), as presented in the following Proposition.
	
	\begin{proposition}
		For any given DPV $\bs{x}$, if the constraints in \eqref{eqn_Collision} are not satisfied, it can always be renumbered (jointly with the precoding matrix $\mbf{W}$) to satisfy \eqref{eqn_Collision}, without degrading the achieved EE performance.
	\end{proposition}
	\begin{IEEEproof}
		See Appendix A.
	\end{IEEEproof}
	
	Proposition 1 indicates that it suffices to solve (P1) without accounting for \eqref{eqn_Collision}, denoted as (P1-relax), followed by an additional renumbering step. As illustrated in Fig.~\ref{Fig_Collision}(b), if we renumber the two destination positions, then inter-MA collision can be avoided, and each MA's movement distance or time can be shortened as well. It is important to note that this procedure is merely a mathematical/algorithmic operation performed during the optimization process, without involving any physical reassignment of actuators, radio frequency (RF) chains, or hardware wiring. However, (P1-relax) is still a non-convex optimization problem due to its fractional objective function involving complex mechanical power consumption. In the next section, we first consider a simpler single-user scenario.
	
	\section{Single-User Scenario}
	In the single-user scenario, the EE performance in \eqref{eqn_EE} can be simplified as
	\begin{align}
		\mrm{EE}_{\mrm{SU}}(\bs{x},\bs{w},v)&=\frac{R(\bs{w},\bs{x})}{E_{\mrm{total}}\left(\bs{w},\bs{x},v\right)}\label{eqn_EE_SU}\\
		&=\frac{\left(T-\tau\right)\log_2\left(1+\left|\bs{w}^H\bs{h}(\bs{x})\right|^2/\sigma^2\right)}{\left(T-\tau\right)\left(\left\|\bs{w}\right\|_2^2+P_s\right)+P_M\left(v\right)\sum_{n=1}^{N}\tau_n},\nonumber
	\end{align}
	where $\bs{w}\ib{C}^{N\times1}$ denotes the transmit beamforming at the BS with $\left\|\bs{w}\right\|_2^2\le P_{\max}$, and the subscripts “$k$” in $R\left(\bs{w},\bs{x}\right)$ and $\bs{w}$ are omitted in the single-user scenario without ambiguity. As such, (P1-relax) reduces to the following problem:
	\begin{subequations}
		\begin{align}
			(\text{P2})\quad\underset{\bs{w},\bs{x},v}{\max}\quad&\mrm{EE}_{\mrm{SU}}(\bs{x},\bs{w},v)\nonumber\\
			\mrm{s.t.}\quad & \left\|\bs{w}\right\|_2^2\le P_{\max},\label{eqn_SU_BF}\\
			& \eqref{eqn_MinDist},\eqref{eqn_C2},\eqref{eqn_C3}. \nonumber
		\end{align}
	\end{subequations}
	However, (P2) is still non-convex and challenging to be solved optimally. To address this, we reveal the hidden monotonicity of the EE performance w.r.t. the MA's moving speed first. \vspace{-8pt}
	
	\subsection{Optimal Moving Velocity}
	To cope with the highly nonlinear expression of $P_M\left(v\right)$ w.r.t. $v$, we introduce the following proposition to capture the relationship between EE and MA moving speed.
	\begin{proposition}
		For any given feasible DPV $\bs{x}$ and the BS's transmit beamforming $\bs{w}$, the EE in \eqref{eqn_EE_SU} monotonically increases with $v$.
	\end{proposition}
	\begin{IEEEproof}
		First, we recast the EE in \eqref{eqn_EE_SU} as a more tractable form:
		\begin{align}
			&\mrm{EE}_{\mrm{SU}}(v)=\frac{\left(T-\tau\right)\log_2\left(1+\left|\bs{w}^H\bs{h}(\bs{x})\right|^2/\sigma^2\right)}{\left(T-\tau\right)\left(\left\|\bs{w}\right\|_2^2+P_s\right)+P_M\left(v\right)\sum_{n=1}^{N}\tau_n},\nonumber\\
			&=\frac{\left(vT-\Delta x\right)\log_2\left(1+\left|\bs{w}^H\bs{h}(\bs{x})\right|^2/\sigma^2\right)}{\left(vT-\Delta x\right)\left(\left\|\bs{w}\right\|_2^2+P_s\right)+P_M\left(v\right)\sum_{n=1}^{N}\left|x_n-x_n^0\right|},\nonumber\\
			&=\frac{\log_2\left(1+\left|\bs{w}^H\bs{h}(\bs{x})\right|^2/\sigma^2\right)}{\left\|\bs{w}\right\|_2^2+P_s+f\left(v\right)\sum_{n=1}^{N}\left|x_n-x_n^0\right|},
		\end{align}
		where $\Delta x\triangleq\underset{n\ic{N}}{\max}\left|x_n-x_n^0\right|$ and $f\left(v\right)\triangleq\frac{P_M\left(v\right)}{vT-\Delta x}$, respectively. It is evident that EE increases with $v$ if $f(v)$ decreases with $v$. The first-order derivative of $f(v)$ is given by
		\begin{equation}
			\frac{\partial f(v)}{\partial v}=\frac{v\frac{\partial M(\omega)}{\partial \omega}\left(vT-\Delta x\right)-l_0M\left(\frac{v}{l_0}\right)\Delta x}{l_0^2\left(vT-\Delta x\right)^2}.
		\end{equation}
		Note that $vT-\Delta x>0$ due to the constraints in \eqref{eqn_Velocity}. Moreover, for a practical stepper motor, its pull-out torque decreases with its angular velocity, as depicted in Fig.~\ref{Fig_Motor}(a). Hence, we have $\frac{\partial M(\omega)}{\partial \omega}<0$. Therefore, we can conclude that $\frac{\partial f(v)}{\partial v}<0$, i.e., $f(v)$ monotonically decreases with $v$. This completes the proof.
	\end{IEEEproof}
	
	Proposition 2 indicates that to maximize the EE of the considered mechanically-driven MA system in the single-user scenario, the stepper motor should always operate at its maximum speed, i.e., $v=v_{\max}$, which also helps prolong the time for data transmission. To corroborate this finding, we plot in Fig.~\ref{Fig_SU_Velocity} the EE performance versus the MA's velocity. The DPV is set as $x_n=x_n^0+\lambda/4$, $\forall n\ic{N}$, and the transmit beamforming vector is set as the maximum-ratio transmission (MRT) based on $\bs{h}(\bs{x})$, i.e., $\bs{w}=\sqrt{P_{\max}}\frac{\bs{h}(\bs{x})}{\left\|\bs{h}(\bs{x})\right\|_2}$. It is observed that the EE performance monotonically increases with the MA's velocity $v$. By substituting $v=v_{\max}$ into (P2), (P2) can be simplified as
	\begin{subequations}
		\begin{align}
			(\text{P3})\quad\underset{\bs{w},\bs{x}}{\max}\quad&\mrm{EE}_{\mrm{SU}}(\bs{w},\bs{x},v_{\max})\nonumber\\
			\mrm{s.t.}\quad &\frac{|x_n-x_n^0|}{T}\le v_{\max},\forall n\ic{N},\label{eqn_MaxDistCons}\\
			&\eqref{eqn_MinDist},\eqref{eqn_C2},\eqref{eqn_SU_BF}.\nonumber
		\end{align}
	\end{subequations}\vspace{-15pt}
	
	\begin{figure}[t]
		\centering
		\captionsetup{justification=raggedright,singlelinecheck=false}
		\centerline{\includegraphics[width=0.45\textwidth]{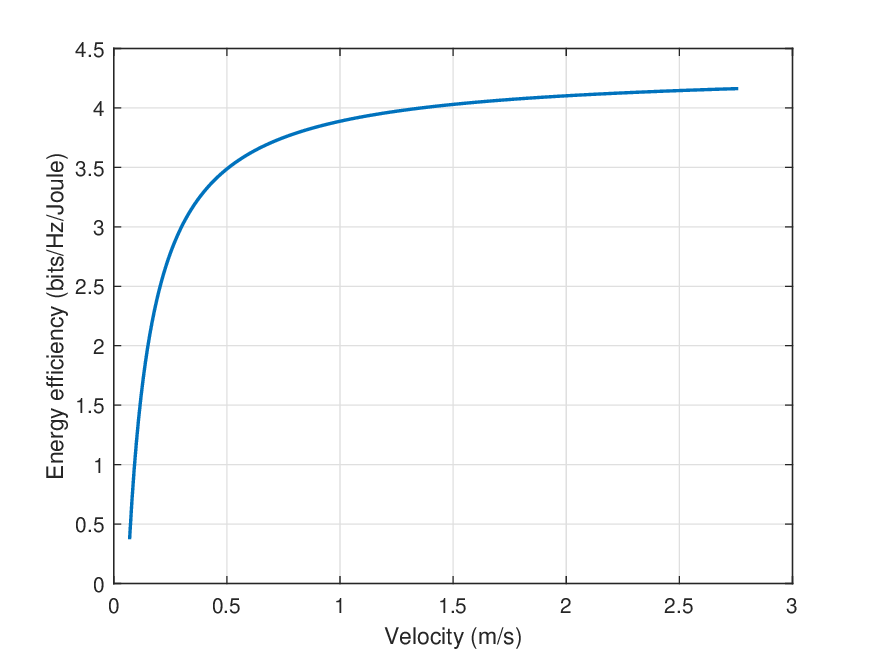}}
		\captionsetup{font=footnotesize}
		\caption{EE performance versus the MA's velocity.}
		\label{Fig_SU_Velocity}
		\vspace{-5pt}
	\end{figure}
	
	\subsection{Proposed Algorithm for (P3)}
	However, (P3) is still difficult to be optimally solved. In the following, we propose a two-layer optimization algorithm to decompose it into two simpler subproblems. The inner problem optimizes the transmit beamforming vector $\bs{w}$ with a given DPV $\bs{x}$, while the outer problem optimizes the DPV $\bs{x}$ based on the optimal value of the inner problem. The inner and outer problems are respectively given by
	\begin{align*}
		(\text{P3-1})\quad\underset{\bs{w}}{\max}\quad\mrm{EE}_{\mrm{SU}}(\bs{w},\bs{x},v_{\max}),\quad\mrm{s.t.}\quad \eqref{eqn_SU_BF}. \nonumber
	\end{align*}
	and
	\begin{align*}
		(\text{P3-2})\,\,\underset{\bs{x}}{\max}\,\,\mrm{EE}_{\mrm{SU}}\left(\bs{w}\left(\bs{x}\right),\bs{x},v_{\max}\right)\quad\mrm{s.t.}\,\,\eqref{eqn_MinDist},\eqref{eqn_C2},\eqref{eqn_MaxDistCons},\nonumber
	\end{align*}
	where $\bs{w}(\bs{x})$ denotes the optimal solution to (P3-1). In the following, we elaborate on the approaches for solving these two subproblems.
	
	\subsubsection{Inner-Layer Optimization}
	To solve (P3-1), we first rewrite $\bs{w}$ into the following form:
	\begin{equation}\label{eqn_SU_BFV}
		\bs{w}=\sqrt{p}\bs{w}_0,
	\end{equation}
	where $p$ denotes the BS's transmit power with $0\le p \le P_{\max}$ and $\bs{w}_0$ denotes the normalized transmit beamforming vector with $\left\|\bs{w}_0\right\|_2^2=1$. By substituting \eqref{eqn_SU_BFV} into the objective function of (P3-1), it is not difficult to see that for any given transmit power $p$, the transmit beamforming vector $\bs{w}_0$ should be the MRT based on $\bs{h}\left(\bs{x}\right)$ to maximize the EE, i.e.,
	\begin{equation}\label{eqn_SU_BFV_OPT}
		\bs{w}_0^{\star}\left(\bs{x}\right)=\frac{\bs{h}\left(\bs{x}\right)}{\left\|\bs{h}\left(\bs{x}\right)\right\|_2}.
	\end{equation}
	With \eqref{eqn_SU_BFV} and \eqref{eqn_SU_BFV_OPT}, we can simplify the expression of EE in (P3-1) and recast it as a power control problem, i.e.,
	\begin{align*}
		(\text{P3-1})\,\,\underset{0 \le p \le P_{\max}}{\max}\,\,\mrm{EE}_{\mrm{SU}}\left(p\right)=\frac{a\log_2(1+p\left\|\bs{h}\left(\bs{x}\right)\right\|_2^2/\sigma^2)}{ap+b},\nonumber
	\end{align*}
	where $a\triangleq T-\tau$ and $b\triangleq aP_s+P_M\left(v_{\max}\right)\sum_{n=1}^{N}\tau_n$, respectively. Note that (P3-1) is a classical fractional programming (FP) problem, for which the Dinkelbach's algorithm can be employed. Specifically, in the $l$-th iteration of the Dinkelbach's algorithm, (P3-1) is transformed into the following subtractive form, i.e.,
	\begin{align*}
		(\text{P3-1-}l)\,\,\underset{0 \le p \le P_{\max}}{\max}\,\,f^{(l)}(p),\nonumber
	\end{align*}
	where
	\begin{equation}\label{eqn_SU_DB_ObjFun}
		f^{(l)}(p)\triangleq a\log_2(1+p\left\|\bs{h}\left(\bs{x}\right)\right\|_2^2/\sigma^2)-\eta^{(l-1)}\left(ap+b\right)
	\end{equation}
	and
	\begin{equation}\label{eqn_Eta}
		\eta^{(l-1)}\triangleq \frac{a\log_2(1+p^{(l-1)}\left\|\bs{h}\left(\bs{x}\right)\right\|_2^2/\sigma^2)}{ap^{(l-1)}+b}
	\end{equation}
	denotes the EE value obtained in the $(l-1)$-th Dinkelbach iteration with $p^{(l-1)}$ denoting the optimal transmit power obtained in this iteration. Note that $f^{(l)}(p)$ is a concave function in $p$. Hence, by setting $\frac{\partial f^{(l)}(p)}{\partial p}=0$, the optimal transmit power that maximizes \eqref{eqn_SU_DB_ObjFun} can be obtained as
	\begin{equation}\label{eqn_SU_Opt_TxPw}
		p^{(l)}\left(\bs{x}\right)=\min\left(\left[\hat{p}^{(l)}\left(\bs{x}\right)\right]^+,P_{\max}\right),
	\end{equation}
	where $\hat{p}^{(l)}\left(\bs{x}\right)$ is given by
	\begin{equation}
		\hat{p}^{(l)}\left(\bs{x}\right)=\frac{1}{\eta^{(l-1)}\ln2}-\frac{\sigma^2}{\left\|\bs{h}\left(\bs{x}\right)\right\|_2^2},
	\end{equation}
	and $\left[z\right]^+=\max(z,0)$. Based on \eqref{eqn_Eta} and \eqref{eqn_SU_Opt_TxPw}, we can compute the value of $\eta^{(l)}$ and proceed to the $(l+1)$-th Dinkelbach iteration. It can be shown that $\eta^{(l)}$ monotonically increases with the iteration and converges to an optimal solution to (P3-1) \cite{xu2013energy}. Let $p^{\star}\left(\bs{x}\right)$ denote the optimal transmit power obtained by the Dinkelbach algorithm. Then, the optimal transmit beamforming vector $\bs{w}^{\star}$ for (P3-1) can be expressed as
	\begin{equation}\label{eqn_SU_OPT_BF}
		\bs{w}^{\star}\left(\bs{x}\right)=\sqrt{p^{\star}\left(\bs{x}\right)}\bs{w}_0^{\star}\left(\bs{x}\right)=\sqrt{p^{\star}\left(\bs{x}\right)}\frac{\bs{h}\left(\bs{x}\right)}{\left\|\bs{h}\left(\bs{x}\right)\right\|_2}.
	\end{equation}
	The overall procedures of the proposed algorithm to solve (P3-1) are summarized in Algorithm 1.
	
	\begin{algorithm}[!t]
		\caption{Proposed Algorithm for Solving (P3-1)}
		\label{alg_SU_BfOpt}
		\begin{algorithmic}[1]
			\STATE Initialize $p^{(0)}=0$ and the convergence accuracy $\epsilon$.
			\STATE Set $l=1$.
			\STATE Initialize $\eta^{(l-1)}$ according to \eqref{eqn_Eta}.
			\REPEAT
				\STATE Set $l=l+1$.
				\STATE Calculate $p^{(l)}$ according to \eqref{eqn_SU_Opt_TxPw}.
				\STATE Calculate $\eta^{(l)}$ according to \eqref{eqn_Eta}.
			\UNTIL{$|\eta^{(l)}-\eta^{(l-1)}|<\epsilon$.}
			\STATE Set $p^{\star}=p^{(l)}$.
			\STATE Output $\bs{w}^{\star}$ with $p^{\star}$ according to \eqref{eqn_SU_OPT_BF} as the optimized solutions to (P3-1).
		\end{algorithmic}
	\end{algorithm}
	
	\subsubsection{Outer-Layer Optimization}
	By employing Algorithm 1, we can obtain the optimal value of (P3-1) with any given DPV $\bs{x}$. As such, (P3-2) can be simplified as
	\begin{align*}
		(\text{P3-2})\,\,\underset{\bs{x}}{\max}\,\,\mrm{EE}_{\mrm{SU}}\left(\bs{w}^{\star}\left(\bs{x}\right),\bs{x},v_{\max}\right)\quad\mrm{s.t.}\,\,\eqref{eqn_MinDist},\eqref{eqn_C2},\eqref{eqn_MaxDistCons}.\nonumber
	\end{align*}
	where $\bs{w}^{\star}\left(\bs{x}\right)$ is the optimal transmit beamforming vector given in $\eqref{eqn_SU_OPT_BF}$. Due to the discrete step size of the MA, we can adopt the sequential update algorithm proposed in \cite{wei2025movable,liu2026general} to solve (P3-2) efficiently. Notably, such a sequential update algorithm has also been utilized in many existing works for antenna position optimization with an intractable objective function \cite{wei2024joint,wang2025antenna} even for continuous antenna movements. Compared to the gradient-based algorithms, the sequential update algorithm is applicable to any problem structure and dispenses with the need for complex gradient calculation. Whereas compared to the heuristic antenna position optimization algorithms such as particle swarm optimization (PSO) \cite{xiao2024multiuser}, the sequential update algorithm ensures local optimality and leads to a lower computational complexity, as will also be demonstrated in Section V via simulation.
	
	Specifically, we sequentially update the position of each MA over multiple rounds, each including $N$ iterations. In the $n$-th iteration, we only optimize the position of the $n$-th MA (i.e., $x_n$), while keeping the positions of the other $(N-1)$ MAs fixed. Let us consider the $n$-th iteration in the $r$-th round and denote by $x_j^{(r)}$ the updated position of the $j$-th MA in this round, $1\le j \le n-1$. Then, the set of all feasible destination positions for optimizing $x_n$ is given by
	\begin{equation}\label{eqn_SU_DPVOpt_Set}
		\begin{aligned}
			\ca{X}_n^{(r)}&=\left\{x\left|x\ic{C}_t,\left|x-x_j^{(r)}\right|\ge D_{\min},\forall1\le j \le n-1,\right.\right.\\
			&\qquad\,\,\,\,\left|x-x_j^{(r-1)}\right|\ge D_{\min},\forall n+1\le j \le N,\\
			&\qquad\,\,\,\left.\left|x-x_n^0\right|\le v_{\max}T\right\},\quad 1<n<N.
		\end{aligned}
	\end{equation}
	In addition, we set $\ca{X}_1^{(r)}=\{x|x\ic{C}_t,|x-x_j^{(r-1)}|\ge D_{\min},\forall2\le j \le N,|x-x_1^0|\le v_{\max}T\}$ and $\ca{X}_N^{(r)}=\{x|x\ic{C}_t,|x-x_j^{(r)}|\ge D_{\min},\forall1\le j \le N-1,|x-x_N^0|\le v_{\max}T\}$. Let $\bs{\hat{x}}=[x_1^{(r)},\cdots,x_{n-1}^{(r)},x_n,x_{n+1}^{(r-1)},\cdots,x_{N}^{(r-1)}]^T$. Then, we can optimize $x_n$ as
	\begin{equation}\label{eqn_OptTxPos_n_r}
		x_n^{(r)}=\arg\underset{x_n\ic{X}_n^{(r)}}{\max}\,\,\mrm{EE}_{\mrm{SU}}\left(\bs{w}^{\star}\left(\bs{\hat{x}}\right),\bs{\hat{x}},v_{\max}\right),
	\end{equation}
	which can be optimally solved via an enumeration within $\ca{X}_n^{(r)}$. Next, we can proceed to update the position of the $(n+1)$-th MA in this round. \vspace{-1pt}
	
	Let $\bs{x}^{\star}=[x_1^\star,x_2^\star,\cdots,x_N^\star]^T$ denote the optimized DPV by the multi-round sequential update. The corresponding optimal transmit beamforming vector can be obtained as $\bs{w}^{\star}\left(\bs{x}^{\star}\right)=\sqrt{p^{\star}\left(\bs{x}^{\star}\right)}\frac{\bs{h}\left(\bs{x}^{\star}\right)}{||\bs{h}\left(\bs{x}^{\star}\right)||_2}$. Finally, to ensure that the optimized DPV satisfies the constraints in \eqref{eqn_Collision} as presented in the proof of Proposition 1, we sort the destination positions in an ascending order as $x_{q_1}^\star<x_{q_2}^\star<\cdots<x_{q_N}^\star$, with permutation indices $q_n\ic{N}$. Then, we set $x_n^\star=x_{q_n}^\star$ and $w_n^\star=w_{q_n}^\star$, $n\ic{N}$, where $w_n^{\star}$ denotes the $n$-th entry of $\bs{w}^{\star}\left(\bs{x}^{\star}\right)$.
	
	\begin{algorithm}[!t]
		\caption{Proposed Algorithm for Solving (P3-2)}
		\label{alg_SU}
		\begin{algorithmic}[1]
			\STATE Initialize $x_n^{(0)}$, $\forall n\ic{N}$ and the convergence accuracy $\epsilon$.
			\STATE Set $r=0$.
			\STATE Calculate $\mrm{EE}_{\mrm{SU}}^{(r)}$ with $x_n=x_n^{(0)}$, $\forall n\ic{N}$ according to \eqref{eqn_EE_SU}.
			\REPEAT
			\STATE Set $r = r+1$.
				\FOR{$n=1\rightarrow N$}
					\STATE Construct $\ca{X}_n^{(r)}$ according to \eqref{eqn_SU_DPVOpt_Set}.
					\STATE Update $x_n^{(r)}$ via \eqref{eqn_OptTxPos_n_r}.
				\ENDFOR
			\UNTIL{the fractional increase of the objective value of (P3-2) is below the given threshold $\epsilon$.}
			\STATE Set $\bs{x}^{\star}=[x_1^\star,x_2^\star,\cdots,x_N^\star]^T$, with $x_n^{\star}=x_n^{(r)}$, $n\ic{N}$.
			\STATE Calculate $\bs{w}^{\star}\left(\bs{x}^{\star}\right)=[w_1^{\star},w_2^{\star},\cdots,w_N^{\star}]^T$ via \eqref{eqn_SU_OPT_BF}.
			\STATE Sort $\bs{x}^{\star}$ in an ascending order as $x_{q_1}^{\star}<x_{q_2}^{\star}<\cdots<x_{q_N}^{\star}$.
			\STATE Set $x_n^\star=x_{q_n}^\star$ and $w_n^\star=w_{q_n}^\star$, $\forall n\ic{N}$.
			\STATE Output $\bs{w}^{\star}\left(\bs{x}^{\star}\right)$ and $\bs{x}^{\star}$ as the optimized solutions to (P3-2).
		\end{algorithmic}
	\end{algorithm}
	
	The overall algorithm for solving (P3-2) is summarized in Algorithm~\ref{alg_SU}. Since each iteration yields a non-decreasing value of (P3-2), the convergence of the proposed two-layer optimization framework is guaranteed. The computational complexity for Algorithm~\ref{alg_SU} is given by $\ca{O}\left(I_{\bs{w}}I_{\bs{x}}NM\right)$, where $I_{\bs{w}}$ denotes the number of iterations of the Dinkelbach algorithm in solving (P3-1), and $I_{\bs{x}}$ denotes the number of sequential update rounds in solving (P3-2).
	
	\section{Multi-User Scenario}
	In this section, we focus on the general multi-user scenario, i.e., (P1-relax). First, we show that the optimal MA moving velocity is still $v_{\max}$ in the multi-user scenario. To this end, we rewrite the EE in \eqref{eqn_EE} into the following form:
	\begin{equation}\label{eqn_EE_MU}
		\begin{aligned}
			\mrm{EE}(\mbf{W},\bs{x},v)&=\frac{(T-\tau)\sum_{k=1}^{K}R_k\left(\mbf{W},\bs{x}\right)}{\sum_{n=1}^{N}{\tau_n P_M\left(v\right)}+\left(T-\tau\right)P_D\left(\mbf{W}\right)}\\
			&=\frac{\sum_{k=1}^{K}R_k\left(\mbf{W},\bs{x}\right)}{f(v)\sum_{n=1}^{N}|x_n-x_n^0|+P_D\left(\mbf{W}\right)},
		\end{aligned}
	\end{equation}
	where $f(v)=\frac{P_M(v)}{vT-\Delta x}$ and $\Delta x=\underset{n\ic{N}}{\max}\,\,|x_n-x_n^0|$. As shown in the proof of Proposition 2, $f(v)$ decreases with $v$. Hence, the EE in \eqref{eqn_EE_MU} in the multi-user scenario also increases with $v$, and (P1-relax) can be simplified as
	\begin{align*}
		\text{(P4)}\quad\underset{\mbf{W},\bs{x}}{\max}\,\,\mrm{EE}\left(\mbf{W},\bs{x},v_{\max}\right)\quad\mrm{s.t.}\,\,\eqref{eqn_MinDist},\eqref{eqn_C1},\eqref{eqn_C2},\eqref{eqn_MaxDistCons},
	\end{align*}
	by setting $v=v_{\max}$. However, (P4) remains difficult to solve due to the strong coupling between the design variables $\mbf{W}$ and $\bs{x}$. To address this difficulty, we develop an AO algorithm to alternately optimize these variables.
	
	\subsection{Proposed AO Algorithm for (P4)}
	\subsubsection{Optimizing $\mbf{W}$ with Given $\bs{x}$}
	First, we optimize the transmit precoding matrix $\mbf{W}$ with any given DPV $\bs{x}$. Under this condition, (P4) can be simplified as
	\begin{align*}
		\text{(P4-1)}\quad\underset{\mbf{W}}{\max}\,\,\mrm{EE}\left(\mbf{W},\bs{x},v_{\max}\right)\quad\mrm{s.t.}\,\,\eqref{eqn_C1}.
	\end{align*}
	Similar to the single-user setup, we apply the Dinkelbach's algorithm to deal with the fractional programming problem. Specifically, in the $l$-th iteration of the Dinkelbach's algorithm, (P4-1) is transformed into the following subtractive form:
	\begin{align*}
		(\text{P4-1-}l)\quad\underset{\mbf{W}}{\max}\quad g^{(l)}\left(\mbf{W}\right)\quad
		\mrm{s.t.}&\quad\eqref{eqn_C1},
	\end{align*}
	where
	\begingroup\makeatletter\def\f@size{9}\check@mathfonts\def\maketag@@@#1{\hbox{\m@th\normalsize\normalfont#1}}%
	\begin{equation}
		g^{(l)}\left(\mbf{W}\right)\triangleq a\sum_{k=1}^{K}R_k\left(\mbf{W},\bs{x}\right)-\eta^{(l-1)} \left(a\mrm{Tr}\left(\mbf{W}\mbf{W}^H\right)+b\right)
	\end{equation}
	\endgroup
	and
	\begin{equation}\label{eqn_Eta_MU}
		\eta^{(l-1)}\triangleq\frac{a\sum_{k=1}^{K}R_k\left(\mbf{W}^{(l-1)},\bs{x}\right)}{a\mrm{Tr}\left(\mbf{W}^{(l-1)}\left(\mbf{W}^{(l-1)}\right)^H\right)+b}
	\end{equation}
	and $\mbf{W}^{(l-1)}$ denotes the optimized transmit beamforming matrix in $(l-1)$-th iteration. However, (P4-1-$l$) is still non-convex due to the complicated expression of $R_k(\mbf{W},\bs{x})$. To proceed, we introduce a slack variable $\bs{\chi}=[\chi_1,\chi_2,\cdots,\chi_K]$ for the SINR term inside $R_k(\mbf{W},\bs{x})$, i.e., $\gamma_k(\mbf{W},\bs{x})$, and transform (P4-1-$l$) into
	\begin{subequations}
		\begin{align}
			(\text{P4-1-}l)\,\,\underset{\mbf{W},\bs{\chi}}{\max}&\quad \bar{g}^{(l)}\left(\mbf{W},\bs{\chi}\right)\nonumber\\
			\mrm{s.t.}&\quad \gamma_k(\mbf{W},\bs{x})\ge \chi_k,\forall k\ic{K},\label{eqn_MU_C1}\\
			&\quad\eqref{eqn_C1},\nonumber
		\end{align}
	\end{subequations}
	where
	\begin{equation}
		\begin{aligned}
			\bar{g}^{(l)}\left(\mbf{W},\bs{\chi}\right)\triangleq &a\sum_{k=1}^{K}\log_2\left(1+\chi_k\right)\\
			&-\eta^{(l-1)} \left(a\mrm{Tr}\left(\mbf{W}\mbf{W}^H\right)+b\right).
		\end{aligned}
	\end{equation}
	To handle the non-convex constraints in \eqref{eqn_MU_C1}, we introduce another positive slack variable $\bs{\xi}=[\xi_1,\xi_2,\cdots,\xi_K]$ and reformulate \eqref{eqn_MU_C1} into an equivalent form:
	\begin{equation}\label{eqn_MU_C1-1}
		\left|\bs{h}_k^H\left(\bs{x}\right)\bs{w}_k\right|^2\ge\xi_k\chi_k,\,\,\forall k,
	\end{equation}
	and
	\begin{equation}\label{eqn_MU_C1-2}
		\sum_{i\ne k}^{K}{\left|\bs{h}_k^H\left(\bs{x}\right)\bs{w}_i\right|^2}+\sigma^2\le \xi_k,\,\,\forall k.
	\end{equation}
	It can be observed that \eqref{eqn_MU_C1-2} is convex while \eqref{eqn_MU_C1-1} still remains non-convex. Nonetheless, the term $\bs{h}_k^H\left(\bs{x}\right)\bs{w}_k$ in constraints \eqref{eqn_MU_C1-1} can be expressed as a real number through an arbitrary rotation to the beamforming vector $\bs{w}_k$. As a result, \eqref{eqn_MU_C1-1} is equivalent to $\Re(\bs{h}_k^H(\bs{x})\bs{w}_k)\ge\sqrt{\xi_k\chi_k}$. As such, by replacing the concave term $\sqrt{\xi_k\chi_k}$ with its first-order Taylor expansion at given feasible points $\{\chi_k^{(t-1)}\}_{k=1}^K$ and $\{\xi_k^{(t-1)}\}_{k=1}^K$, the constraints in \eqref{eqn_MU_C1-1} can be expressed as
	\begin{align}\label{eqn_MU_C1-1-Approx}
		\Re(\bs{h}_k^H(\bs{x})\bs{w}_k)\ge&\sqrt{\xi_k^{(t-1)}\chi_k^{(t-1)}}+\frac{1}{2}\sqrt{\frac{\chi_k^{(t-1)}}{\xi_k^{(t-1)}}}\left(\xi_k-\xi_k^{(t-1)}\right)\nonumber\\
		&+\frac{1}{2}\sqrt{\frac{\xi_k^{(t-1)}}{\chi_k^{(t-1)}}}\left(\chi_k-\chi_k^{(t-1)}\right),\,\,\forall k.
	\end{align}
	Based on the above approximations, (P4-1-$l$) can be formulated approximately as the following problem
	\begin{subequations}
		\begin{align}
			(\text{P4-1-}l\text{-}t)\quad\underset{\mbf{W},\bs{\chi},\bs{\xi}}{\max}&\quad \bar{g}^{(l)}\left(\mbf{W},\bs{\chi}\right)\nonumber\\
			\mrm{s.t.}&\quad\eqref{eqn_C1},\eqref{eqn_MU_C1-2},\eqref{eqn_MU_C1-1-Approx},\nonumber\\
			&\quad\xi_k > 0,\forall k\ic{K},
		\end{align}
	\end{subequations}
	which is convex and can be efficiently solved by the interior-point method. The overall algorithm to solve (P4-1) is summarized in Algorithm~\ref{alg_MU_BfOpt}.
	
	\begin{algorithm}[!t]
		\caption{Proposed Algorithm for Solving (P4-1)}
		\label{alg_MU_BfOpt}
		\begin{algorithmic}[1]
			\STATE Initialize $\mbf{W}^{(0)}$ and the convergence accuracy $\epsilon_1$.
			\STATE Set $l=0$.
			\STATE Initialize $\eta^{(l)}$ according to \eqref{eqn_Eta_MU}.
			\REPEAT
				\STATE Set $l=l+1$.
				\STATE Initialize $\bs{\chi}^{(0)}$, $\bs{\xi}^{(0)}$, and convergence accuracy $\epsilon_2$.
				\STATE Set $t=1$.
				\REPEAT
					\STATE Solve (P4-1-$l$-$t$) with given $\bs{\chi}^{(t-1)}$ and $\bs{\xi}^{(t-1)}$ via the interior-point method.
					\STATE Update $\boldsymbol{\chi}^{(t)}$ and $\boldsymbol{\xi}^{(t)}$ according to the solution of (P4-1-$l$-$t$).
					\STATE Set $t=t+1$.
				\UNTIL{the fractional increase of the objective value of (P4-1-$l$) is below the given threshold $\epsilon_2$.}
				\STATE Output $\mbf{W}^{(l)}$ as the solution to (P4-1-$l$).
				\STATE Calculate $\eta^{(l)}$ according to \eqref{eqn_Eta_MU}.
			\UNTIL{$\left|\eta^{(l)}-\eta^{(l-1)}\right|\le \epsilon_1$}
			\STATE Output $\mbf{W}^{(l)}$ as the optimized solutions to (P4-1).
		\end{algorithmic}
	\end{algorithm}
	
	\subsubsection{Optimizing $\bs{x}$ with Given $\mbf{W}$}
	Next, we optimize the DPV $\bs{x}$ with any given transmit precoding matrix $\mbf{W}$, which is equivalent to solving the following problem:
	\begin{align*}
		\text{(P4-2)}\quad\underset{\bs{x}}{\max}\,\,\mrm{EE}\left(\mbf{W},\bs{x},v_{\max}\right)\quad\mrm{s.t.}\,\,\eqref{eqn_MinDist},\eqref{eqn_C2},\eqref{eqn_MaxDistCons}.
	\end{align*}
	Notably, (P4-2) shares a similar structure to (P3-2), differing primarily in the objective function. Therefore, the multi-round sequential update algorithm used to solve (P3-2) can also be applied here, by changing the objective function in (P3-2) to $\mrm{EE}\left(\mbf{W},\bs{x},v_{\max}\right)$ with a given transmit precoding matrix $\mbf{W}$. For brevity, the details are omitted.
	
	Let $\bs{x}^{\star} = [x_1^{\star}, x_2^{\star}, \cdots, x_N^{\star}]^T$ denote the optimized (but unsorted) DPV obtained by the above process in each AO iteration. Similar to the single-user scenario, we sort $\bs{x}^{\star}$ in an ascending order as $x_{q_1}^{\star}<x_{q_2}^{\star}<\cdots<x_{q_N}^{\star}$ and set $x_n^{\star}=x_{q_n}^{\star}$, thereby avoiding inter-antenna collisions. In addition, the optimized transmit precoding matrix in Section IV-A1 should be updated as well, i.e., $\mbf{W}\left[n,:\right]=\mbf{W}\left[q_n,:\right]$, where $\mbf{W}[n,:]$ denotes the $n$-th row of $\mbf{W}$.
	
	\subsubsection{Overall Algorithm}
	Based on the above, we can alternately solve (P4-1) and (P4-2) via Algorithms~\ref{alg_MU_BfOpt} and \ref{alg_SU}, respectively. The overall algorithm to solve (P4) is summarized in Algorithm~\ref{alg_MU}. Note that each iteration of the SCA procedure in Algorithm~\ref{alg_MU_BfOpt} produces a non-decreasing objective value, ensuring that (P4-1) converges to its maximum. Similarly, Algorithm~\ref{alg_SU} yields a non-decreasing objective value of (P4-2). Hence, the overall convergence of Algorithm~\ref{alg_MU} is guaranteed.
	
	Finally, we analyze the computational complexity of Algorithm~\ref{alg_MU}. According to \cite{wang2014outage}, the computational complexity for solving (P4-1) is given by $\ca{O}\left(I_{\bs{w}}N^{6}K^{1.5}\right)$. In addition, the computational complexity for solving (P4-2) is given by $\ca{O}\left(I_{\bs{x}}NM\right)$. As a result, the total complexity of Algorithm~\ref{alg_MU} is given by $\ca{O}\left(I_{\bs{w}}N^{6}K^{1.5}+I_{\bs{x}}NM\right)$.
	
	\begin{algorithm}[!t]
		\caption{Proposed Algorithm for Solving (P4)}
		\label{alg_MU}
		\begin{algorithmic}[1]
			\STATE Initialize $\mbf{W}^{(0)}$, $\bs{x}^{(0)}$, the convergence accuracy $\epsilon$.
			\STATE Calculate the EE performance $\mrm{EE}^{(0)}$ with $\mbf{W}^{(0)}$ and $\bs{x}^{(0)}$. Set $l=0$.
			\REPEAT
			\STATE Set $l=l+1$.
			\STATE Obtain $\mbf{W}^{(l)}$ with given $\bs{x}^{(l-1)}$ according to Algorithm~\ref{alg_MU_BfOpt}.
			\STATE Obtain $\bs{x}^{(l)}$ with given $\mbf{W}^{(l)}$ via a similar process to Algorithm~\ref{alg_SU}.
			\STATE Sort $\bs{x}^{(l)}$ in an ascending order as $x_{q_1}^{(l)}<x_{q_2}^{(l)}<\cdots<x_{q_N}^{(l)}$.
			\STATE Update $x_n^{(l)}=x_{q_n}^{(l)}$ and $\mbf{W}^{(l)}\left[n,:\right]=\mbf{W}^{(l)}\left[q_n,:\right]$, $n\ic{N}$.
			\STATE Calculate the EE performance $\mrm{EE}^{(l)}$ with $\mbf{W}^{(l)}$ and $\bs{x}^{(l)}$.
			\UNTIL{$\left|\mrm{EE}^{(l)}-\mrm{EE}^{(l-1)}\right|\le \epsilon$}
			\STATE Output $\mbf{W}^{(l)}$ and $\bs{x}^{(l)}$ as the optimized solutions to (P4).
		\end{algorithmic}
	\end{algorithm}
	
	\section{Numerical Results}
	In this section, we present numerical results to validate the efficacy of our proposed algorithms. Unless otherwise specified, the simulation parameters are set as follows. The carrier wavelength is $\lambda=0.06$ meter (m). The length of the linear array is $A=6\lambda=0.36$ m. The initial positions of the $N$ MAs are symmetrically distributed along the center of $\ca{C}_t$ with a half-wavelength spacing. The number of MAs is $N=6$, and that of users is $K=2$. The BS's static circuit power consumption is $P_s=30$ dBm, and its maximum transmit power is $P_{\max}=30$ dBm. The time duration of a channel coherence block is $T=0.25$ s. We consider the field-response-based channel model presented in \cite{zhu2025tutorial} for MAs in the simulation. The distance from the BS to each user is uniformly distributed between 20 m and 100 m, and the path-loss exponent for the BS-user channels is $\alpha=2.8$. The number of multipath components in the BS-user channels is $L=10$, and their path gains are assumed to follow the CSCG distribution, i.e., $g_{k,l}\sim\ca{CN}(0,\rho d_k^{-\alpha}/L)$, $l=1,2,\cdots,L$, $k\ic{K}$, where $\rho$ represents the path loss at the reference distance of 1 m. The angles of departure (AoDs) for these paths are assumed to be independent and identically distributed variables following the uniform distribution within $[-\frac{\pi}{2},\frac{\pi}{2}]$. The average noise power is $\sigma^2=-80$ dBm. Moreover, we consider the AM2224 high-speed stepper motor in this simulation \cite{am2224}, with the same parameters as those adopted in Fig.~\ref{Fig_Motor}. The maximum angular speed of the stepper motor is $\omega_{\max}=552$ rad/s. The step angle and radius of the lead screw are $\omega_{D}=\frac{\pi}{12}$ rad and $l_0=5$ mm, respectively. Hence, the step size of the stepper motor is $d_s=\omega_{D}l_0\approx1.2$ mm. All the results are averaged over 1000 independent channel realizations.
	
	Furthermore, we consider the following benchmark schemes for performance comparison:
	\begin{enumerate}
		\item \textbf{Benchmark 1: PSO.} In this benchmark, the DPV $\bs{x}$ is optimized via the PSO method in the continuous space \cite{xiao2024multiuser} and then quantized based on the step size $d_s$, and the transmit precoding matrix $\mbf{W}$ is optimized via a similar process as proposed in this paper.
		\item \textbf{Benchmark 2: Conventional EE optimization (ConvEE).} The optimization problem (P1) is first solved by neglecting the mechanical power consumption, i.e., $P_M(v)=0$, via our proposed algorithms. Then, the actual EE performance involving the mechanical power consumption is calculated with the optimized MA positions and transmit precoding.
		\item \textbf{Benchmark 3: Sum-rate maximization (SM).} In this benchmark, the transmit precoding matrix $\mbf{W}$ and DPV $\bs{x}$ are optimized to solely maximize the achievable sum-rate $\sum_{k=1}^{K}R_k(\mbf{W},\bs{x})$. The transmit precoding matrix is optimized via the weighted minimum mean square error (WMMSE) algorithm with a fixed DPV $\bs{x}$ \cite{shi2011iteratively}; while the DPV $\bs{x}$ is optimized via the sequential update method presented in Algorithm~\ref{alg_SU}.
		\item \textbf{Benchmark 4: FPA.} In this benchmark, the DPV is fixed as $x_n=x_n^0$, and the BS's transmit precoding matrix is optimized via the Dinkelbach's algorithm.
	\end{enumerate}
	
	\subsection{Single-User System}
	\begin{figure}[t]
		\centering
		\captionsetup{justification=raggedright,singlelinecheck=false}
		\centerline{\includegraphics[width=0.45\textwidth]{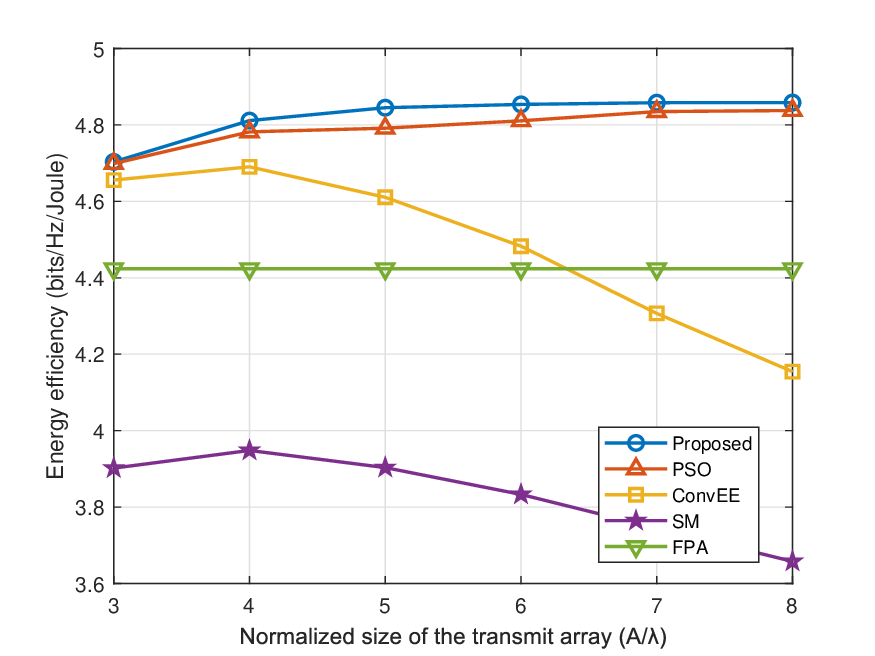}}
		\captionsetup{font=footnotesize}
		\caption{EE performance versus the normalized size of transmit array.}
		\label{Fig_SU_TxRegion}
		\vspace{-15pt}
	\end{figure}
	\begin{figure}[t]
		\centering
		\captionsetup{justification=raggedright,singlelinecheck=false}
		\centerline{\includegraphics[width=0.45\textwidth]{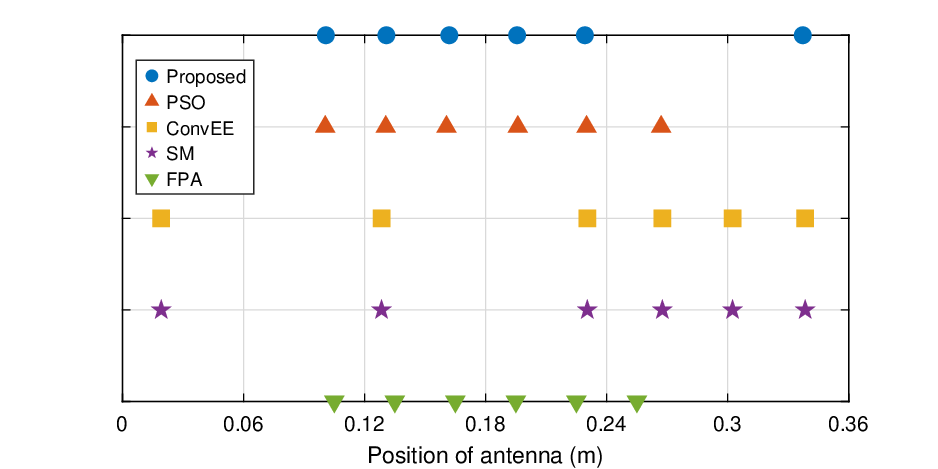}}
		\captionsetup{font=footnotesize}
		\caption{Optimized positions of the MAs by different schemes.}
		\label{Fig_SU_OptPos}
		\vspace{-15pt}
	\end{figure}
	\begin{figure}[t]
		\centering
		\captionsetup{justification=raggedright,singlelinecheck=false}
		\centerline{\includegraphics[width=0.45\textwidth]{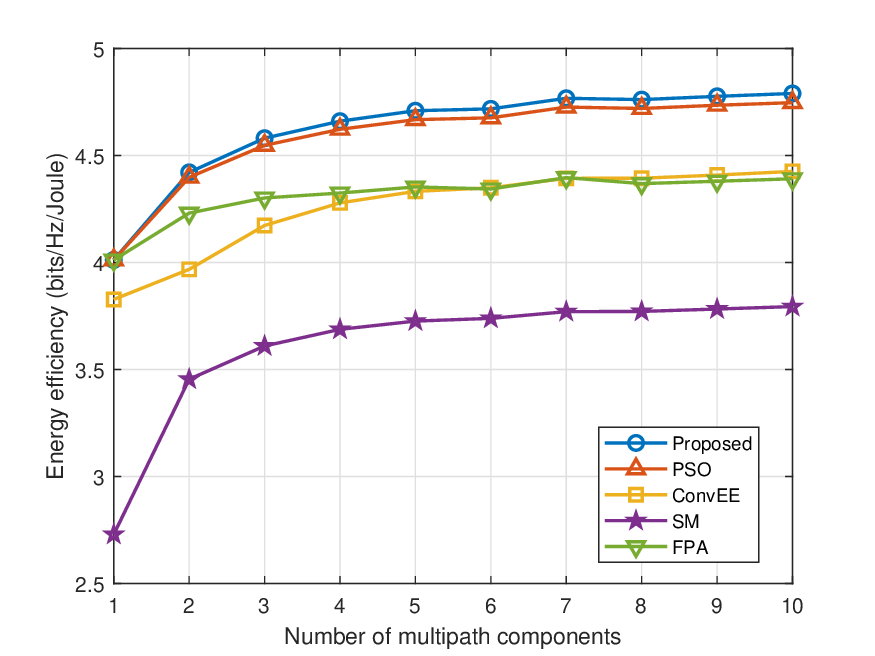}}
		\captionsetup{font=footnotesize}
		\caption{EE performance versus the number of multipath components.}
		\label{Fig_SU_PathNum}
		\vspace{-15pt}
	\end{figure}
	\begin{figure}[t]
		\centering
		\captionsetup{justification=raggedright,singlelinecheck=false}
		\centerline{\includegraphics[width=0.45\textwidth]{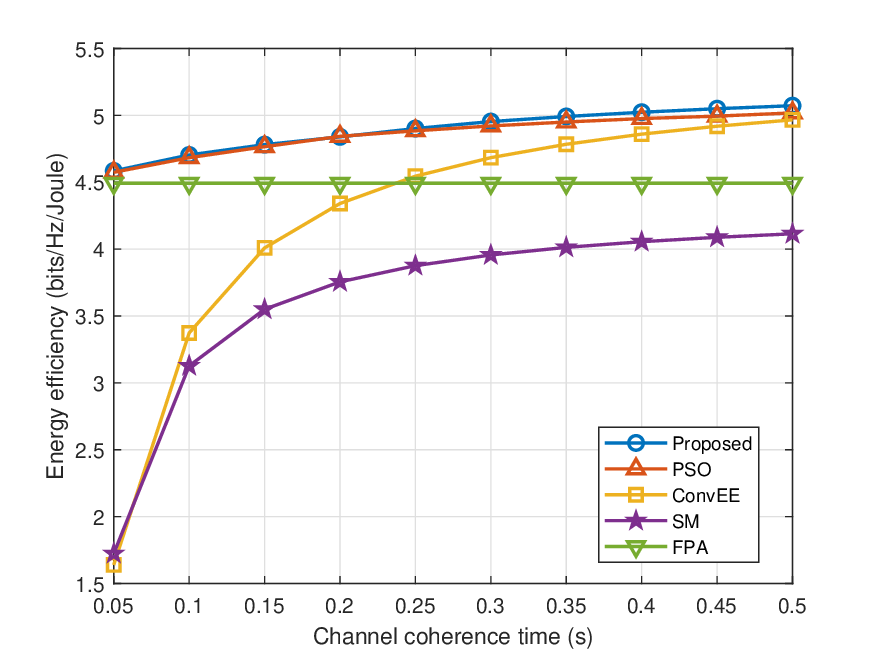}}
		\captionsetup{font=footnotesize}
		\caption{EE performance versus the channel coherence time.}
		\label{Fig_SU_TotalTime}
		\vspace{-15pt}
	\end{figure}
	First, we consider the single-user scenario and plot in Fig.~\ref{Fig_SU_TxRegion} the EE performance versus the normalized size of the transmit array (i.e., $A/\lambda$). It is observed that the EE performance of our proposed algorithm increases with the size of $\ca{C}_t$ and outperforms the PSO algorithm and other benchmarks. In contrast, the ConvEE benchmark exhibits a decline in EE performance as the size of $\ca{C}_t$ increases, due to its omission of mechanical power consumption in optimization. As a result, the optimized DPV leads to longer movement distances from their associated CPVs, thus degrading the EE performance. This highlights the necessity of accounting for mechanical power consumption in mechanically-driven MA systems. Furthermore, the EE performance of the SM benchmark is observed to be significantly worse compared to the other schemes. This behavior results from its exclusive focus on rate maximization while disregarding both mechanical and transmit power consumption, leading to substantial overall power consumption and degraded EE performance.

	\begin{figure}[t]
		\centering
		\captionsetup{justification=raggedright,singlelinecheck=false}
		\centerline{\includegraphics[width=0.45\textwidth]{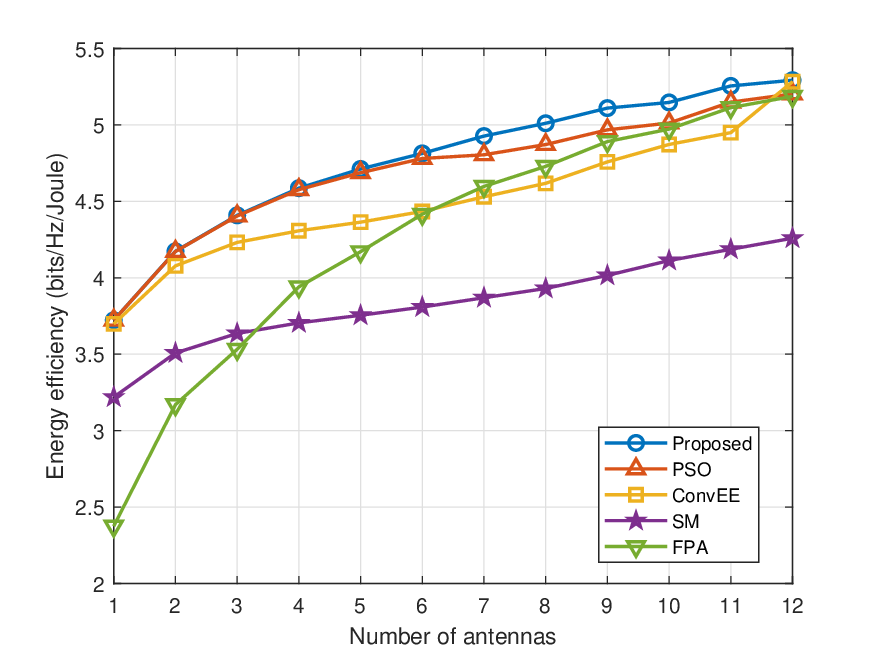}}
		\captionsetup{font=footnotesize}
		\caption{EE performance versus the number of antennas.}
		\label{Fig_SU_TxNums}
		\vspace{-15pt}
	\end{figure}
	
	To gain more insights, we plot in Fig.~\ref{Fig_SU_OptPos} the optimized positions of the MAs by different schemes in one specific channel realization, with $N=6$ and $A=6\lambda$. It is observed that the optimized MA positions by the proposed algorithm are located closer to the initial positions (i.e., those of the FPA scheme) compared to those by the ConvEE and SM benchmarks, thereby reducing the mechanical energy consumption and movement delay due to the long-distance movement. In contrast, the optimized positions by the PSO algorithm are observed to be even closer to the initial antenna positions compared to the proposed algorithm. Although this proximity helps reduce the mechanical power consumption, it also leads to less effective exploitation of the spatial diversity within the movement region. As such, the PSO algorithm needs to consume more transmit power to compensate for the loss in channel power gain, which thus reduces its EE performance. It follows that the proposed algorithm yields a better balance between the rate performance and power consumption than the other benchmark schemes.
	
	Next, we plot in Fig.~\ref{Fig_SU_PathNum} the EE performance versus the number of multipath components in the BS-user channel (i.e., $L$). It is observed that the EE performance of all schemes (except the FPA benchmark) improves as the number of paths increases. This is because when $L$ increases, the channel power gain within the movable region exhibits more significant fluctuation, resulting in enhanced spatial diversity gain. This helps shorten the movement time to identify the optimal antenna position that balances rate performance and energy consumption. In addition, the EE performance of all schemes except the SM benchmark is observed to be identical for $L=1$. This is because under the MRT beamforming strategy in \eqref{eqn_SU_BFV_OPT}, the channel power gain $||\bs{h}(\bs{x})||_2^2$ becomes constant and independent of antenna position.  It is also observed that the ConvEE and SM benchmarks yield the worst EE performance among all considered schemes, as similarly observed from previous figures.
	
	Fig.~\ref{Fig_SU_TotalTime} depicts the EE performance versus the channel coherence time $T$ with $N=6$. It is observed that the EE performance of all schemes (except the FPA benchmark) improves with increasing $T$, as this increases each MA's maximum allowable moving distance as seen from \eqref{eqn_Velocity}. Hence, more favorable antenna positions can be explored to improve the rate-energy trade-off. In contrast, the EE performance of the FPA benchmark keeps constant as $T$ increases. Furthermore, it is observed that the improvement in the EE performance by the proposed scheme decreases with $T$ and ultimately converges. This is because as $T$ is sufficiently large, the EE in \eqref{eqn_EE} will degrade to the conventional EE presented in \eqref{eqn_EE_infty}, which is regardless of $T$, as discussed at the end of Section II. This also results in decreased performance gap between the proposed algorithm and the ConvEE benchmark as $T$ increases.
	
	Lastly, we plot in Fig.~\ref{Fig_SU_TxNums} the EE performance versus the number of antennas (i.e., $N$). It is observed that the EE performance of all considered schemes monotonically increases with $N$. On one hand, this is due to the improved beamforming gain by increasing $N$. On the other hand, this is due to the reduced movement distance for each MA to cover as $N$ increases considering the finite region size. It is also observed that the performance gaps between the proposed algorithm and the FPA and ConvEE benchmarks reduce as $N$ increases. This is because there exists a maximum allowable number of MAs given the distance constraints in \eqref{eqn_MinDist}, i.e., $N_{\max}=A/D_{\min} = 12$. Thus, when $N$ approaches 12, the flexibility of the antenna position optimization reduces, thus resulting in less performance gains from antenna movement.
	
	\subsection{Multi-User System}
	\begin{figure}[t]
		\centering
		\captionsetup{justification=raggedright,singlelinecheck=false}
		\centerline{\includegraphics[width=0.45\textwidth]{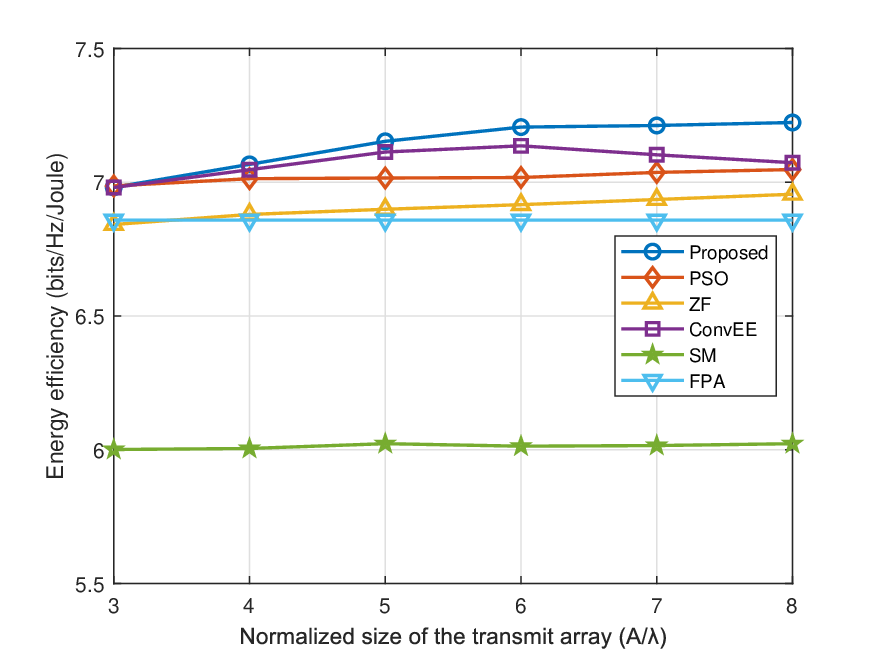}}
		\captionsetup{font=footnotesize}
		\caption{EE performance versus the normalized size of transmit region.}
		\label{Fig_MU_TxRegion}
		\vspace{-15pt}
	\end{figure}
	
	In this subsection, we present the simulation results for the multi-user scenario. Besides the baseline schemes adopted in the single-user scenario, we include a benchmark based on zero-forcing (ZF) precoding, where $\mbf{W}$ is fixed as the ZF precoder. The resulting power allocation and DPV $\bs{x}$ can be optimized in a two-layer manner similarly to Algorithm~\ref{alg_SU}. Specifically, in the inner layer, the power allocation is optimized via the Dinkelbach algorithm for any given $\bs{x}$ \cite{huang2019reconfigurable}. In the outer layer, the DPV $\bs{x}$ is optimized via the sequential update algorithm as in Section III-B2. This benchmark is marked as “ZF” in the subsequent figures.
	
	First, we plot in Fig.~\ref{Fig_MU_TxRegion} the EE performance versus the normalized size of the transmit array, i.e., $A/\lambda$. It is observed that the proposed algorithm consistently outperforms all benchmark schemes, while the ZF benchmark slightly outperforms the FPA benchmark. Moreover, compared with the single-user scenario as shown in Fig.~\ref{Fig_SU_TxRegion}, the EE improvement by increasing $A$ is less significant in the multi-user setup. This is primarily due to the fact that in the single-user scenario, the MAs can be positioned with full flexibility to maximize the EE for that user. In contrast, in the multi-user case, the optimized MA positions need to jointly balance the desired and interference channels for multiple users, as well as the associated mechanical power consumption, which limits the achievable EE gains. Nonetheless, when $A=8\lambda$, the proposed algorithm can still achieve an approximately 0.5 bps/Hz/Joule improvement over the FPA benchmark. It is also observed that both our proposed algorithm and the ZF benchmark can ensure a monotonic increase in EE as $A$ increases. In contrast, the EE performance of the ConvEE benchmark degrades as $A \ge 6\lambda$, while that of the SM benchmark keeps almost constant and remains the worst among all considered schemes. This highlights the importance of balancing the data rate and mechanical energy consumption in the multi-user scenario as well.
	
	\begin{figure}[t]
		\centering
		\captionsetup{justification=raggedright,singlelinecheck=false}
		\centerline{\includegraphics[width=0.45\textwidth]{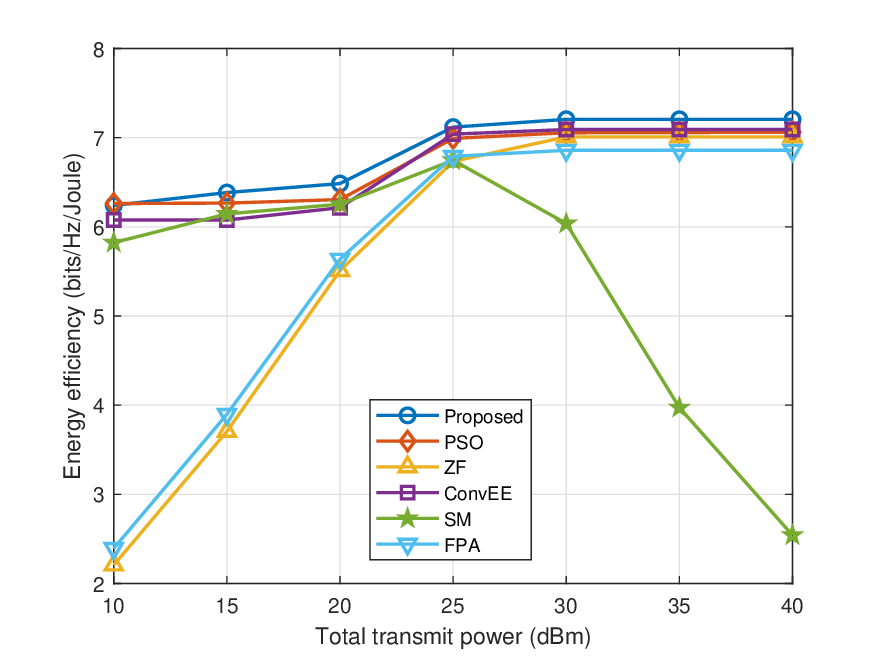}}
		\captionsetup{font=footnotesize}
		\caption{EE performance versus the maximum transmit power.}
		\label{Fig_MU_TxPower}
		\vspace{-15pt}
	\end{figure}
	
	Next, we plot in Fig.~\ref{Fig_MU_TxPower} the EE performance versus the maximum transmit power $P_{\max}$ in the multi-user scenario. It is observed that the EE performance of all considered schemes improves with $P_{\max}$ when $P_{\max}\le25$ dBm. This is because the EE performance is more dominated by the achievable rate in the low transmit power regime, compared to the total power consumption. Hence, all schemes should transmit at the maximum power to maximize the achievable rate. It is also interesting to note that unlike conventional EE maximization in FPA systems, there still exists certain performance gap (around 0.4 bps/Hz/Joule) between the proposed algorithm and SM benchmark in the low transmit power regime. This is because the additional mechanical power consumption also contributes to the EE performance, and the SM benchmark may introduce more significant mechanical power consumption compared to the proposed method. In addition, a significant performance gap between the proposed method and the FPA benchmark is observed in the low transmit power regime, implying that the rate gain from antenna movement can be dominant over the resulting mechanical power consumption for EE maximization in this regime. The ZF benchmark is observed to achieve an even worse performance than FPA as $P_{\max} \le 15$ dBm. This is because, in the low-power regime, the received SINR of each user is primarily limited by the noise power $\sigma^2$ instead of inter-user interference. Consequently, the ZF benchmark, which focuses on interference suppression, may offer limited benefits.
	
	On the other hand, when $P_{\max} > 25$ dBm, the total power consumption plays a more significant role. As a result, transmitting at the maximum power may result in a significant loss in EE, as observed from the SM benchmark. In contrast, the performance of other schemes is observed to remain constant as $P_{\max}$ increases, suggesting that their transmit powers are unchanged to maximize the EE. Furthermore, the ZF benchmark is observed to achieve a close performance to the proposed method in the high transmit power regime, due to the more prominent effects of inter-user interference on the EE performance.
	
	\begin{figure}[t]
		\centering
		\captionsetup{justification=raggedright,singlelinecheck=false}
		\centerline{\includegraphics[width=0.45\textwidth]{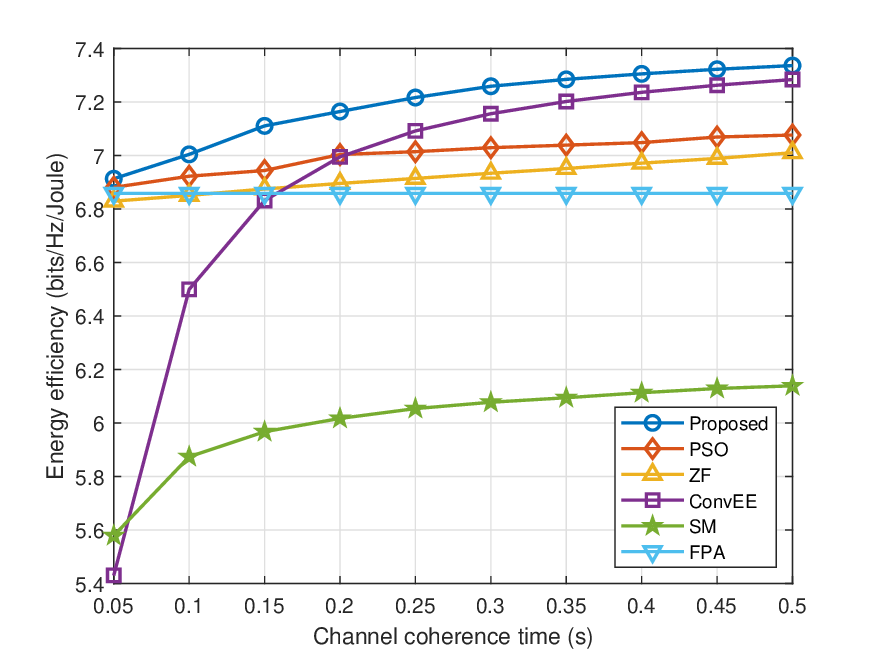}}
		\captionsetup{font=footnotesize}
		\caption{EE performance versus the channel coherence time.}
		\label{Fig_MU_TotalTime}
		\vspace{-15pt}
	\end{figure}
	
	In Fig.~\ref{Fig_MU_TotalTime}, we plot the EE performance versus the channel coherence time (i.e., $T$). Similar to the observations made from Fig.~\ref{Fig_SU_TotalTime}, the EE performance by all schemes increases with $T$, and our proposed algorithm achieves the highest EE performance among them. In addition, the performance gap between our proposed algorithm and the ConvEE benchmark decreases with $T$, due to the less significant effects of mechanical power consumption. However, their performance gap amounts to 1.5 bps/Hz/Joule as $T=0.05$ s. It is also observed that the performance gap between our proposed algorithm and the FPA benchmark increases with $T$ due to the enlarged data transmission time. The ZF benchmark is also observed to outperform the FPA benchmark when $T>0.1$ s.
	
	\begin{figure}[t]
		\centering
		\captionsetup{justification=raggedright,singlelinecheck=false}
		\centerline{\includegraphics[width=0.45\textwidth]{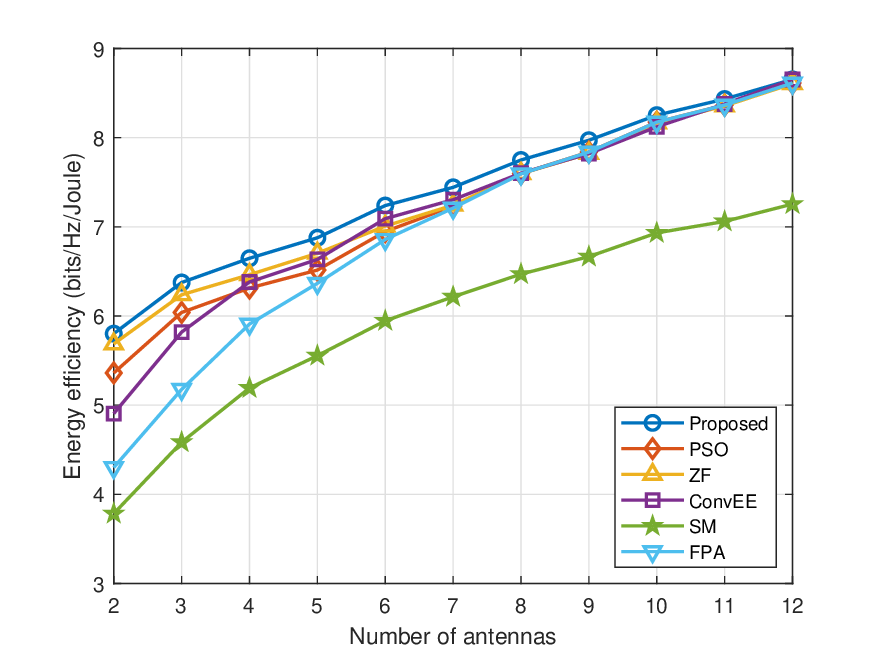}}
		\captionsetup{font=footnotesize}
		\caption{EE performance versus the number of antennas.}
		\label{Fig_MU_TxNums}
		\vspace{-15pt}
	\end{figure}
	
	Lastly, we plot in Fig.~\ref{Fig_MU_TxNums} the EE performance versus the number of antennas, i.e., $N$. It is observed that the performance of all considered schemes increases with $N$, and our proposed algorithm consistently yields a higher EE performance compared to the other schemes, especially for small values of $N$. For example, when $N=2$, the proposed scheme can yield an EE improvement over the ConvEE and SM benchmarks by 0.9 bps/Hz/Joule and 1.6 bps/Hz/Joule, respectively. However, the performance advantage of our proposed scheme diminishes as $N$ increases. This is because the maximum allowable number of MAs is bounded by $A/D_{\min}$; hence, the movable region for each individual antenna shrinks significantly as $N$ increases. This phenomenon also occurs for electronically driven MAs, since the same inter-antenna spacing constraints need to be enforced to avoid mutual coupling.
	
	
	\section{Conclusions and Future Work}
	In this paper, we investigated an EE maximization problem for a stepper motor-driven multi-MA system. First, we integrated a classical power consumption model for stepper motors into the conventional energy consumption model of MA systems. Based on this overall model, we formulated an EE maximization problem by jointly optimizing the MAs' destination positions, moving speed, and the BS's transmit precoding matrix, subject to the inter-MA collision avoidance constraints in their movement. To address this challenging optimization problem, we revealed that the inter-MA collision avoidance constraints can be safely relaxed and all MAs should move at their maximum velocity to maximize the EE. For the remaining BS precoding and MA destination position optimization problem, we proposed a two-layer optimization framework and an AO algorithm to solve it in the single- and multi-user scenarios, respectively.
	
	Numerical results are provided to show the efficacy of our proposed algorithms in boosting the EE by properly balancing the rate performance and overall energy consumption. The main takeaways are summarized as follows. First, compared to conventional FPAs, MAs can explore more spatial diversity to improve the EE, especially in the low transmit power regime. Second, compared to the SM scheme, our proposed algorithm can avoid long-range MA movement that incurs substantial energy consumption and degrades EE. Third, unlike FPA systems, rate maximization may only achieve suboptimal performance even in the low transmit power regime in MA systems. Fourth, EE maximization without accounting for mechanical energy consumption can lead to a significant EE loss in the cases of low channel coherence time, large antenna movement region, and small BS antenna number. Last but not least, the EE performance gain by MAs becomes less pronounced in the multi-user scenario compared to the single-user scenario, which is consistent with the observations made from achievable rate maximization with MAs \cite{zhu2024enhanced,wei2025movable,zhang2025sum,wu2024globally}.
	
	This paper can be extended to various directions as future work. First, it would be interesting to investigate the EE maximization for a more general 2D planar array, which introduces additional spatial degrees of freedom but a more challenging collision avoidance and moving trajectory design, for which the renumbering procedure is not directly applicable. Second, this paper assumes that each MA is driven via a dedicated stepper motor. It is worthy of studying the EE characterization and maximization under other MA architectures, such as cross-linked MAs, 6DMAs with antenna rotation, array-level MAs, extremely large MAs, among others. Third, EE maximization under other system setups can also be studied by accounting for mechanical power consumption, such as joint BS- and user-side MAs, physical-layer security, ISAC, mobile computing, and so on. Fourth, this paper assumes perfect position control based on high-precision stepper motors, while practical implementations may encounter dynamic errors such as actuator latency or mechanical drift. It is thus interesting to develop more robust optimization algorithms to explicitly account for these position errors.\vspace{-8pt}
	
	\appendices
	\section{Proof of Proposition 1}
	Let $\bs{x}=\left[x_1,x_2,\cdots,x_N\right]^T\ib{R}^{N\times1}$ and $\mbf{W}=\left[\bs{w}_1,\bs{w}_2,\cdots,\bs{w}_K\right]\ib{C}^{N\times K}$ denote the optimized DPV and transmit precoding matrix without the collision-avoidance constraints in \eqref{eqn_Collision}, respectively. We sort $\bs{x}$ in an ascending order as $x_{q_1}<x_{q_2}<\cdots<x_{q_N}$ with permutation indices $q_n\ic{N}$. Then, we renumber $\bs{x}$ and $\mbf{W}$ as $x_n=x_{q_n}$ and $\bs{w}_n=\bs{w}_{q_n}$, $n\ic{N}$, respectively. It has been proven in \cite[\textbf{Theorem 1}]{li2025trajectory} that the resulting trajectory from the CPV to the DPV $\bs{x}$ is collision-free.
	
	Let $\tau_n=\frac{|x_n-x_n^0|}{v}$ and $\tau_n'=\frac{|x_{q_n}-x_n^0|}{v}$ denote the movement delay of the $n$-th MA before/after the above renumbering, respectively. Next, we prove that this renumbering will not decrease the EE performance. Specifically, we rewrite the expression of EE in \eqref{eqn_EE} into a more tractable form as
	\begin{equation}\label{eqn_Appendix_Eq1}
		\begin{aligned}
			\mrm{EE}(\mbf{W},\bs{x},v)&=\frac{(T-\tau)\sum_{k=1}^{K}R_k\left(\mbf{W},\bs{x}\right)}{E_{\mrm{total}}\left(\mbf{W},\bs{x},v\right)}\\
			&=\frac{(T-\tau)\sum_{k=1}^{K}R_k\left(\mbf{W},\bs{x}\right)}{\sum_{n=1}^{N}{\tau_n P_M\left(v\right)}+\left(T-\tau\right)P_D\left(\mbf{W}\right)}\\
			&=\frac{\sum_{k=1}^{K}R_k\left(\mbf{W},\bs{x}\right)}{P_M\left(v\right)\frac{\sum_{n=1}^{N}\tau_n}{T-\tau}+P_D\left(\mbf{W}\right)}.
		\end{aligned}
	\end{equation}
	It is straightforward to see that the renumbering procedure does not affect the values of the sum-rate $\sum_{k=1}^{K}R_k\left(\mbf{W},\bs{x}\right)$, the total transmit power $P_D\left(\mbf{W}\right)$, and the power consumption for the MA driver $P_M(v)$. In addition, as shown in \cite[\textbf{Theorem 1}]{li2025trajectory}, the maximum delay after the renumbering is no greater than that before the renumbering, i.e.,
	\begin{equation}
		\tau'=\underset{n\ic{N}}{\max}\,\,{\tau_n'}\le\tau=\underset{n\ic{N}}{\max}\,\,{\tau_n},
	\end{equation}
	or equivalently, $T-\tau' \ge T-\tau$. Hence, if
	\begin{equation}\label{eqn_Appendix_TotDelayInequality}
		\sum_{n=1}^{N}\tau_n'\le\sum_{n=1}^{N}\tau_n,
	\end{equation}
	then the EE performance after the renumbering must be no worse than that before the renumbering. To prove \eqref{eqn_Appendix_TotDelayInequality}, we present the following lemma.
	
	\begin{lemma}
		Consider four arbitrary positive numbers $a$, $b$, $c$, and $d$, with $a<b$ and $c<d$. Then, we have
		\begin{equation}
			|a-c|+|b-d|\le|a-d|+|b-c|.
		\end{equation}
	\end{lemma}
	
	Lemma 1 can be proven readily by enumerating all possible relationships among the four numbers. Then, given the DPV $\bs{x}=[x_1,x_2,\cdots,x_N]$, we define a pair $(i,j)$ as a \textit{reverse pair} if $i<j$ but $x_{i}>x_j$, $i,j\ic{N}$. It is evident to see that if there is no reverse pair for the DPV $\bs{x}$, then the equality in \eqref{eqn_Appendix_TotDelayInequality} should hold. If there exists at least one reverse pair $(i,j)$, by applying Lemma 1, we have
	\begin{equation}\label{eqn_Appendix_Eq2}
		\begin{aligned}
		\sum_{n=1}^{N}\tau_n&=\sum_{m\ne i,j}{\tau_m}+\frac{|x_i^0-x_i|}{v}+\frac{|x_j^0-x_j|}{v}\\
			&\ge\sum_{m\ne i,j}{\tau_m}+\frac{|x_i^0-x_j|}{v}+\frac{|x_j^0-x_i|}{v},
		\end{aligned}
	\end{equation}
	which indicates that the total delay can be reduced by swapping the indices $i$ and $j$. Since the number of reverse pairs is finite, the swapping process in \eqref{eqn_Appendix_Eq2} can be repeated for all remaining reverse pairs until no reverse pairs remain, which leads to $x_n=x_{q_n}$, $\forall n\ic{N}$. Hence, the inequality in \eqref{eqn_Appendix_TotDelayInequality} is established. Based on the above, \eqref{eqn_Appendix_TotDelayInequality} must hold after the renumbering. Hence, Proposition 1 is proved.
	
	\bibliography{MA_EE.bib}
	\bibliographystyle{IEEEtran}
\end{document}